\documentclass[lettersize,journal]{IEEEtran}
\usepackage{amsthm}
\usepackage{cite}
\usepackage{amsmath,amsfonts}
\usepackage{graphicx}
\usepackage{textcomp}
\usepackage{xcolor}
\usepackage{CJKutf8}
\usepackage{tabularray}
\usepackage{subfigure}
\usepackage{caption}
\usepackage{multirow}
\usepackage{bm}
\usepackage{amsmath,amsfonts}
\usepackage{array}
\usepackage{subcaption}
\usepackage[caption=false,font=normalsize,labelfont=sf,textfont=sf]{subfig}
\usepackage{textcomp}
\usepackage{stfloats}
\usepackage{url} 
\usepackage{verbatim}
\usepackage{graphicx}
\usepackage{cite}
\newtheorem{theorem}{Theorem}
 \usepackage{subcaption}
\usepackage{amsmath,amsfonts}
\usepackage{array}
\usepackage[caption=false,font=normalsize,labelfont=sf,textfont=sf]{subfig}
\usepackage{textcomp}
\usepackage{stfloats}
\usepackage{url}
\usepackage{verbatim}
\usepackage{graphicx}
\def\BibTeX{{\rm B\kern-.05em{\sc i\kern-.025em b}\kern-.08em
    T\kern-.1667em\lower.7ex\hbox{E}\kern-.125emX}}
\usepackage{balance}
\usepackage{amsmath}
\usepackage{subcaption}
\allowdisplaybreaks[4]
\usepackage{multicol}
\usepackage{cuted}
\usepackage{multicol,amsmath}
\usepackage{makecell}
\usepackage{array}
\newcolumntype{C}{>{\centering\arraybackslash}X}
\usepackage[ruled,lined,linesnumbered]{algorithm2e}
\usepackage{tabularx,booktabs}
\usepackage[flushleft]{threeparttable}

\usepackage{booktabs,ragged2e}
\usepackage{multirow}
\usepackage{subcaption}
\usepackage{bm}
\begin{document}

\allowdisplaybreaks[4]
\allowdisplaybreaks[4]

\begin{CJK}{UTF8}{gbsn}

\title{A Superposition Code-Based Semantic Communication Approach with Quantifiable and Controllable Security}


\author{Weixuan Chen,~\IEEEmembership{Graduate Student Member, IEEE}, Shuo Shao,~\IEEEmembership{Member, IEEE}, Qianqian Yang,~\IEEEmembership{Member, IEEE}, \\ Zhaoyang Zhang,~\IEEEmembership{Senior Member, IEEE}, and Ping Zhang,~\IEEEmembership{Fellow, IEEE}
        
\thanks{
W. Chen, Q. Yang$^{\dag}$, and Z. Zhang are with the College of Information Science and Electronic Engineering, Zhejiang University, Hangzhou,
China. (e-mails: \{12231075, qianqianyang20$^{\dag}$, ning\_ming\}@zju.edu.cn).

S. Shao is with the School of Cyber Science and Engineering, Shanghai Jiao Tong University, Shanghai, China. (e-mail: shuoshao@sjtu.edu.cn). 

P. Zhang is with the State Key Laboratory
of Networking and Switching Technology, Beijing University of Posts and
Telecommunications, Beijing, China. (e-mail: pzhang@bupt.edu.cn).

This work is partly supported by the NSFC under grant No. 62293481, No. 62201505, by the National Key R\&D Program of China under Grant 2024YFE0200802, and by the Zhejiang Provincial Natural Science Foundation of China under Grant No. LZ25F010001.  (Corresponding author: Qianqian Yang.)  
}
}

\maketitle

\begin{abstract}
This paper addresses the challenge of achieving security in semantic communication (SemCom) over a wiretap channel, where a legitimate receiver coexists with an eavesdropper experiencing a poorer channel condition. 
Despite previous efforts to secure SemCom against eavesdroppers, 
guarantee of approximately zero information leakage remains an open issue. 
In this work, we propose a secure SemCom approach based on superposition code, aiming to provide quantifiable and controllable security for digital SemCom systems. 
The proposed method employs a double-layered constellation map, where semantic information is associated with satellite constellation points and cloud center constellation points are randomly selected. By carefully allocating power between these two layers of constellation, we ensure that the symbol error probability (SEP) of the eavesdropper when decoding satellite constellation points is nearly equivalent to random guessing, while maintaining a low SEP for the legitimate receiver to successfully decode the semantic information. 
%
Simulation results demonstrate that the peak signal-to-noise ratio (PSNR) and mean squared error (MSE) of the eavesdropper's reconstructed data, under the proposed method, can range from decoding Gaussian-distributed random noise to approaching the variance of the data. 
This validates the effectiveness of our method in nearly achieving the experimental upper bound of security for digital SemCom systems when both eavesdroppers and legitimate users utilize identical decoding schemes.
Furthermore, the proposed method consistently outperforms benchmark techniques, showcasing superior data security and robustness against eavesdropping.
The implementation code is publicly available at: \url{https://github.com/1weixuanchen/A-Superposition-Code-Based-Semantic-Communication}.

\end{abstract}

\begin{IEEEkeywords}
Digital semantic communications, superposition code, wiretap channel, upper bound of security.
\end{IEEEkeywords}



\section{Introduction}
Semantic communication (SemCom) \cite{qin2021semantic,10766058}, emerging as a novel communication paradigm, has received significant attention in recent years due to its capability to facilitate efficient transmission for data-intensive applications. 
This innovative approach focuses on the extraction and transmission of essential information, referred to as \textit{semantic information}\cite{gunduz2022beyond}, crucial for data reconstruction or task execution, while discarding irrelevant details. 
Consequently, SemCom stands out as a bandwidth-efficient communication paradigm that enhances the transmission efficiency of communication systems and improves the quality of intelligent information services. 
Deep learning (DL) techniques, proven highly effective in various domains \cite{he2016deep,otter2020survey}, play a pivotal role in SemCom systems by leveraging neural networks (NNs) to perform joint source and channel coding (Deep JSCC) \cite{bourtsoulatze2019deep} on the source data. This integration of DL enables SemCom systems to achieve excellent performance in transmitting various types of information, such as text\cite{farsad2018deep,han2022semantic}, speech\cite{weng2023deep,han2022semanticpre}, image \cite{lee2019deep,jankowski2020wireless,zhang2022semantic}, video\cite{jiang2022wireless}, and multimodal data \cite{xie2022task}.

The progress in SemCom also raises concerns regarding security and privacy \cite{shen2023secure,chen2023model}.
Erdemir \textit{et al.} \cite{erdemir2022privacy} proposed a secure SemCom system designed for wiretap channels.
This approach utilized a variational autoencoder (VAE)-based NN architecture and incorporated a specially designed loss function for end-to-end training, striking a balance between minimizing information leakage to potential eavesdropper and maintaining low distortion at the legitimate receiver. In a related work, Marchioro \textit{et al.} \cite{marchioro2020adversarial} developed a data-driven secure SemCom scheme leveraging an generative adversarial training method by treating the eavesdropper as an adversary and penalizing the information leakage. Note that both works assume knowledge of the eavesdropper's network to assist in training the legitimate user's network. Specifically, it is assumed that the legitimate user can minimize the correlation between the eavesdropper's output and the classification label to prevent information leakage. However, this assumption may be impractical in real-world scenarios, as eavesdroppers typically do not collaborate with legitimate users. Moreover, treating information leakage as a penalty in the training process, rather than strictly constraining it, does not fundamentally secure the SemCom system.

Different from the aforementioned schemes, Tung \textit{et al.} \cite{tung2023deep} utilized a conventional public key cryptographic scheme proposed in \cite{lindner2011better} to encrypt the signals to be transmitted, thus protecting the proposed Deep JSCC scheme for wireless image transmission against eavesdroppers. Notably, this approach alleviates the assumption of knowing the eavesdropper's model. However, it still falls short of ensuring the security of SemCom systems due to its vulnerability to quantum computing attacks \cite{buchanan2017will,yan2013quantum} and traditional attack methods, including related-key attacks \cite{biham1994new}. Furthermore, similar to the aforementioned works, it focuses on analog SemCom systems, posing challenges for deployment in the current wireless communication systems, as the majority of practical communication systems rely on digital communication.

To address these challenges, this paper explores an approach to nearly achieving zero information leakage for DL-based digital SemCom systems over a wiretap channel \cite{wyner1975wire} with additive white Gaussian noise (AWGN). Similarly, the considered system consists of a transmitter referred to as \textit{Alice}, a legitimate receiver called \textit{Bob}, and an eavesdropper called \textit{Eve}. 
Alice aims to transmit the semantic information  extracted from source data $\textbf{X}$ to Bob over an AWGN channel. However, Eve can wiretap the transmitted data and attempt to reconstruct the source data over another AWGN channel with worse channel condition, i.e., larger noise. The objective of this paper is to minimize the distortion in reconstructing $\textbf{X}$ at Bob end while ensuring that the information leakage to Eve approaches zero.

However, developing such a secure approach faces two main challenges.
The signals sent by Alice contain semantically significant information extracted by NN-based encoders for the purpose of source reconstruction. This information can be captured by Eve over wiretap channel, enabling partial recovery of the source data.  
Additionally, traditional capacity-achieving coding schemes for Gaussian wiretap channel \cite{tyagi2014explicit} utilize a superposition code with two layers, employing a codebook derived through a complex two-step random codebook generation process. The intricate nature of this process makes it difficult for NN-based encoders to learn.

Inspired by the traditional coding scheme, we propose a superposition code based secure digital SemCom approach, involving three main steps. 
Firstly, we impose two constellation maps in digital modulation \cite{bo2024jcm} to implement the superposition code. For instance, by overlaying the constellation map of 4-quadrature amplitude modulation (4-QAM) modulation onto another 4-QAM constellation map, we create a 16-QAM constellation map. The symbol sequence of the inner constellation map is considered the cloud center codeword of the superposition code, while that of the outer constellation map is the satellite codeword. Secondly, we modulate the extracted semantic information onto the outer constellation map and randomly generate symbols for the inner constellation map. This randomly generated inner constellation point is inspired by the classic coding scheme in the wiretap channel and serves as a random key for encrypting the semantic information.
Last, we dynamically adjust the power allocation between the inner and outer constellation maps based on the symbol error probability (SEP) of both the legitimate receiver and the eavesdropper. 
The optimal power allocation is determined by minimizing the SEP of the legitimate receiver while maintaining a constraint on the SEP of the eavesdropper. 
The numerical results demonstrate that the proposed method reduces the mutual information between the eavesdropper's reconstructed data and the source data to nearly zero when both eavesdroppers and legitimate users utilize identical decoding schemes, thereby ensuring approximately zero information leakage. 
This is achieved even under high compression ratios, providing a strong guarantee of system security. In addition, we achieve controllable and quantifiable security by adjusting the SEP constraints imposed on eavesdroppers.


\section{Related Work}

In this section, we review related works on SemCom with superposition coding, as well as the application of superposition coding in enhancing communication security.
We also review the security and privacy aspects of SemCom. 
This allows readers to develop a comprehensive understanding of the field and reflects the distinctiveness of our work.

%

Regarding SemCom with superposition coding,  
Li \textit{et al.} \cite{li2023nonortho} proposed a non-orthogonal multiple access (NOMA)-enhanced multi-user SemCom system, where superposition coding allows the simultaneous transmission of different semantic data to multiple users, significantly improving spectral efficiency. 
Bo \textit{et al.} \cite{bo2024deep} introduced deep learning-based superposition coded modulation (DeepSCM) for hierarchical SemCom, where different levels of semantic information are encoded and superposed for efficient broadcast.
%
These approaches demonstrate that superposition coding is a viable technique for integration into SemCom systems, and they motivate our exploration of its application in different SemCom contexts.

Regarding superposition coding for enhancing communication security \cite{csiszar2003broadcast,el2011network},
Xu \textit{et al.} \cite{xu2021quantum} used nonrandom superimposed coding to secure pilot signals in 5G URLLC systems. By superimposing multiple pilot signals and applying quantum learning to identify and eliminate attack-induced uncertainties, this approach ensures that pilot signals are reliably decoded, even despite pilot-aware attacks, thus securing the uplink access phase.
%
Xu \textit{et al.} \cite{xu2019secure} demonstrated that superposition coding can enhance physical layer security in sparse mmWave massive MIMO systems. 
By mapping confidential signals onto the dominant angular components within a layered transmission framework, their method leverages spatial domain characteristics to induce asymmetry between legitimate users and eavesdroppers. This structured signal superposition, aligned with channel sparsity, introduces uncertainty for unauthorized receivers and leads to a measurable gain in secrecy rate.
Tian \textit{et al.} \cite{tian2017secrecy} studied the secrecy sum rate optimization problem for downlink MIMO-NOMA systems, where superposition coding serves as the fundamental mechanism enabling non-orthogonal transmission and enhancing physical layer security.
By exploiting the layered structure of superposition coding, their method enforces a specific decoding order among legitimate users through successive interference cancellation (SIC), ensuring correct signal separation among users.
Moreover, the joint design of precoding within this superposition framework enables the transmitted signal to be deliberately shaped in a way that restricts the eavesdropper's ability to decode confidential information, thereby effectively improving the system's overall secrecy performance.
Additionally,
Xu \textit{et al.} \cite{xu2018high} developed a multi-user covert communication system for avionics, which combines superposition coding with noise modulation to enhance transmission covertness.
While their method focuses on undetectability rather than secrecy, it demonstrates the utility of superposition coding for enhancing the privacy of wireless transmissions.
These works demonstrate that superposition coding provides a natural and effective means of enhancing communication security, and further underscore its potential as a promising approach for securing SemCom systems.
%

Regarding security and privacy protection of SemCom systems, 
Luo \textit{et al.} \cite{luo2023encrypted} proposed an encrypted SemCom system that employs symmetric key encryption and adversarial training to protect privacy in SemCom over wiretap channels. Their design enables both encrypted and unencrypted transmission modes while aiming to resist eavesdropping and maintain communication performance.
Li \textit{et al.} \cite{li2024secure} proposed a secure SemCom system over wiretap channels, leveraging physical layer security. They designed a DNN-based architecture called DeepSSC, and introduced a two-phase training strategy to balance semantic reliability and security by minimizing the semantic information leakage to the eavesdropper.
%
%
These works focus on securing semantic information transmission over wiretap channels and are closely related to our research. 
However, they assume that the architecture and parameters of Eve's model are known, and they do not provide explicit control over the system's security. 
In contrast, our proposed approach requires no prior knowledge of Eve's model and enables quantifiable and controllable security. 
This represents an important yet underexplored direction in secure SemCom, and our work is among the first to study this issue.

\section{Problem Setup and System Design}

\subsection{Problem Setup}
\begin{figure*}[htbp]

\begin{center}
\centerline{\includegraphics[width=1\linewidth]{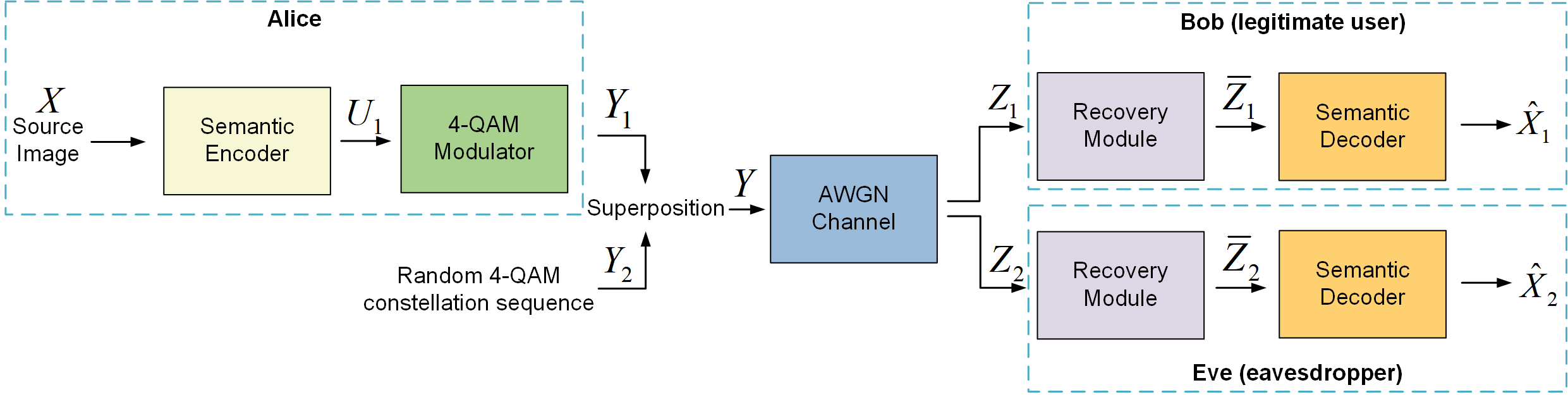}}
\caption{The framework of the proposed secure digital SemCom system.}
\label{fig1}
\end{center}
\vskip -0.2in
\end{figure*}

We consider a digital SemCom system for wireless image transmission over a wiretap channel with AWGN. In this system, a transmitter wants to reliably transmit a source image to a receiver. There also exists a eavesdropper attempting to recover the source image by capturing the noisy transmitted symbols. Fig.~\ref{fig1} shows the framework of our proposed secure digital SemCom system.

We denote the source image to transmit by $\textbf{X}$, the legitimate sender by Alice, the legitimate receiver (also called the legitimate user in this paper) by Bob, and the eavesdropper by Eve, as shown in Fig.~\ref{fig1}.
Alice uses a semantic encoder to extract the semantic information $\textbf{U}_{1}$ from $\textbf{X}$, denoted by
\begin{equation}
\textbf{U}_{1} = f_{\mathrm{se}}\left( {\textbf{X};\bm{\theta}^{\mathrm{se}}} \right),
\end{equation}
where $f_{\mathrm{se}}$ represents the NN-based semantic encoder, and  $\bm{\theta}^{\mathrm{se}}$ refers to the learnable parameters of $f_{\mathrm{se}}$.

Semantic information $\textbf{U}_{1}$ is then input into an NN-based 4-QAM modulator to generate the outer constellation sequence $\textbf{Y}_{1}$, denoted by 
\begin{equation}
\textbf{Y}_{1} = f_{\mathrm{mod}}\left( {\textbf{U}_{1};\bm{\theta}^{\mathrm{mod}}} \right),
\end{equation}
where $f_{\mathrm{mod}}$ represents the 4-QAM modulator, and  $\bm{\theta}^{\mathrm{mod}}$ refers to the learnable parameters of $f_{\mathrm{mod}}$. We then randomly generate another 4-QAM constellation sequence with the same length as $\textbf{Y}_{1}$ as the inner constellation sequence $\textbf{Y}_{2}$. We note that $\textbf{Y}_{2}$ does not carry any valid information. Here, we consider 4-QAM modulation for both the outer and inner constellation for simplicity. 
However, it can be easily extended to higher-order modulation schemes, and this aspect will be discussed in detail in Section V.

We then scale and combine these two 4-QAM constellation sequences to form a 16-QAM constellation sequence $\textbf{Y}$, which is transmitted over a wiretap channel with AWGN.
The superposition operation is controlled by a power allocation coefficient (PAC) $a \in (0,0.5)$, denoted by 
\begin{equation}
\textbf{Y} = \sqrt{a} \cdot \textbf{Y}_1 + \sqrt{1 - a} \cdot \textbf{Y}_2.
\end{equation}
$\textbf{Y}$ is transmitted over the AWGN channel to Bob, who receives a noisy constellation sequence 
\begin{equation}
  \textbf{Z}_{1} = \textbf{Y} + \textbf{n}_1,   
\end{equation}
where $\textbf{n}_1 \sim \mathcal{CN} (0, {\sigma_{1}^{2}} )$ denotes the independent and identically distributed complex Gaussian noise with a mean of 0 and a variance of $\sigma_{1}^{2}$.  Eve eavesdrops through his/her own AWGN channel and receives another noisy constellation sequence
\begin{equation}
\textbf{Z}_{2} = \textbf{Y} + \textbf{n}_2, 
\end{equation}
where $\textbf{n}_2 \sim \mathcal{CN} (0, {\sigma_{2}^{2}} )$.
Since the channel signal-to-noise ratio (SNR) of the legitimate user is usually higher than that of the eavesdropper's receiver, we assume $\sigma_1 < \sigma_2$.

The noisy constellation sequences $\textbf{Z}_{1}$ and $\textbf{Z}_{2}$ are distinct, with $\textbf{Z}_{1}$ corresponding to the legitimate user and $\textbf{Z}_{2}$ to the eavesdropper. 
The key difference lies in the noise levels, as ${\sigma_{1}}$ (for the legitimate user) is smaller than ${\sigma_{2}}$ (for the eavesdropper), meaning that $\textbf{Z}_{2}$ experiences more noise interference than $\textbf{Z}_{1}$.


The channel SNR between Alice and the legitimate user is given by 
\begin{equation}
   \mathrm{SNR}_{\mathrm{leg}} = 10 \log_{10}\left(\frac{P}{\sigma_{1}^{2}}\right) \, (\mathrm{dB}),
\end{equation}
and the channel SNR between Alice and the eavesdropper by 
\begin{equation}
    \mathrm{SNR}_{\mathrm{eve}} = 10 \log_{10}\left(\frac{P}{\sigma_{2}^{2}}\right) \, (\mathrm{dB}).
\end{equation}
Both Bob and Eve share the common objective of recovering the source image. 
Their primary goal is to decode the outer constellation sequence, as the inner constellation sequence contains no valid information. 
Therefore, before being fed into the semantic decoder, both Bob and Eve try to recover the outer constellation sequence from $\textbf{Z}_{1}$ and $\textbf{Z}_{2}$.
Specifically, denote the coordinates of a received constellation point by $(x,y)$, we have
\begin{equation}
    \left( \bar{x},\bar{y} \right) = \left\{ \begin{matrix}
{\left( {x - \sqrt{\frac{1 - a}{2}},y - \sqrt{\frac{1 - a}{2}}} \right),x \geq 0,y \geq 0.} \\
{\left( {x + \sqrt{\frac{1 - a}{2}},y - \sqrt{\frac{1 - a}{2}}} \right),x < 0,y \geq 0.} \\
{\left( {x - \sqrt{\frac{1 - a}{2}},y + \sqrt{\frac{1 - a}{2}}} \right),x \geq 0,y < 0.} \\
{\left( {x + \sqrt{\frac{1 - a}{2}},y + \sqrt{\frac{1 - a}{2}}} \right),x < 0,y < 0.}
\end{matrix} \right.,
\end{equation}
where $\left( \bar{x},\bar{y} \right)$ is the recovered outer constellation point from $(x,y)$. Applying this to every point in $\textbf{Z}_{1}$ and $\textbf{Z}_{2}$, we then obtain the recovered outer constellation sequences at Bob and Eve, denoted by $\bar{\textbf{Z}}_{1}$ and $\bar{\textbf{Z}}_{2}$.

For the legitimate user, its semantic decoder performs image recovery based on $\bar{\textbf{Z}}_{1}$, denoted by
\begin{equation}
\hat{\textbf{X}}_{1} = f_{\mathrm{sd}}\left( {\bar{\textbf{Z}}_{1};\bm{\theta}_{1}^{\mathrm{sd}}} \right),
\end{equation}
where $\hat{\textbf{X}}_{1}$ represents the recovered image by the legitimate user, and $f_{\mathrm{sd}}$ represents the semantic decoder of the legitimate user.
Similarly, for the eavesdropper, we have
\begin{equation}
\hat{\textbf{X}}_{2} = g_{\mathrm{sd}}\left( {\bar{\textbf{Z}}_{2};\bm{\theta}_{2}^{\mathrm{sd}}} \right),
\end{equation}
where $\hat{\textbf{X}}_{2}$ represents the recovered image by the eavesdropper, and $g_{\mathrm{sd}}$ represents the semantic decoder of the eavesdropper.

We utilize the peak signal-to-noise ratio (PSNR) as a metric to evaluate the image recovery quality of the users. PSNR is defined by the formula:
\begin{equation}
PSNR(\textbf{X},\hat{\textbf{X}}) = 10\log_{10}\left( \frac{{MAX}^{2}}{MSE(\textbf{X},\hat{\textbf{X}})} \right) (\mathrm{dB}).
\end{equation}
Here, $MAX$ represents the maximum pixel value within the source image, which is 255 for a 24-bit RGB image. 
The $MSE$ is calculated as:
\begin{equation}
MSE(\textbf{X}, \hat{\textbf{X}}) = ||\textbf{X} - \hat{\textbf{X}}||_2^2.
\end{equation}

In addition to pixel-level analysis, we evaluate the perceptual quality of the recovered images using the learned perceptual image patch similarity (LPIPS) metric \cite{zhang2018unreasonableLPIPS}, which is based on the AlexNet architecture.
LPIPS measures the semantic differences between the recovered and source images by comparing deep features extracted from a pre-trained neural network, aligning with human perception. Lower LPIPS values indicate higher perceptual fidelity.

The proposed system aims to ensure that privacy leakage to eavesdroppers meets a given constraint (to be discussed later) while maximizing the performance of legitimate users.

\subsection{Proposed SemCom System}
Our system can be divided into five parts, including semantic encoder, 4-QAM modulator, inner constellation sequence generation, superposition constellation map design, and semantic decoder. The semantic encoder extracts the semantic information, which is then modulated by the 4-QAM modulator into the outer constellation sequence. 
The inner constellation sequence is generated randomly.
Then we generate a superposition constellation sequence combining outer and inner constellation sequences in order to prevent information leakage to the eavesdropper. 
This part will be discussed in detail in Section IV. 
The semantic decoder at the receiver decodes the received signals to generate the recovered image. 
In this paper, we use convolutional neural networks (CNNs) for the architectures of the semantic encoder and decoder due to their well-established efficacy in handling spatial dependencies in image data. 
In fact, our proposed method can be applied to various neural network architectures, and the CNN architecture we used represents one of the most suitable choices.
We describe each part of our system in detail in the following.

\subsubsection{Semantic Encoder and 4-QAM Modulator}
\begin{figure}[htbp]
\begin{center}
\centerline{\includegraphics[width=1\linewidth]{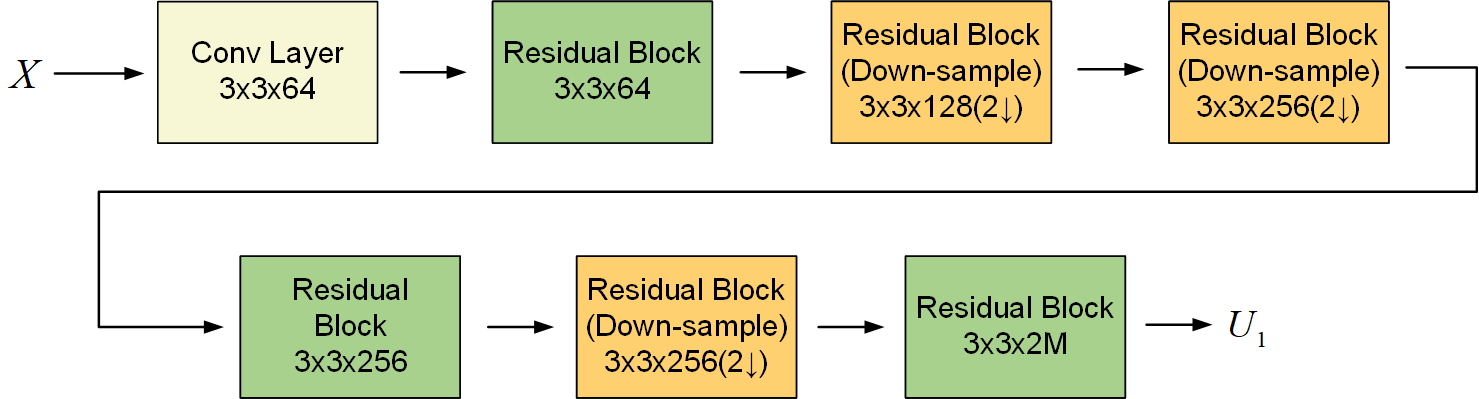}}
\caption{The structure of the semantic encoder.}
\label{fig.2}
\end{center}
\vskip -0.2in
\end{figure}

The network structure of the semantic encoder is shown in Fig.~\ref{fig.2}. 
The semantic encoder is used to extract information from the source image $\textbf{X} \in \mathbb{R}^{H \times W \times 3}$ that is relevant to the image recovery task, i.e., the semantic information $\textbf{U}_{1}$.
Here $H$ and $W$ are the height and width of the source image $\textbf{X}$, respectively. 
The last dimension of $\textbf{X}$ corresponds to the RGB channels.
The semantic encoder consists of one convolutional layer and six residual blocks as shown in Fig.~\ref{fig.2}. The number $ S_{\rm kernel} \times S_{\rm kernel} \times C_{\rm out}$ below each convolutional layer or residual block represents its configuration, where $C_{\rm out}$ is the number of output channels, and $S_{\rm kernel}$ is the kernel size. We note that for the residual block, $S_{\rm kernel}$ is the kernel size of the convolutional layer on the main path. $ 2 \downarrow $ means down-sampling with a stride of 2.
These residual blocks perform feature extraction and down-sampling of the source image $\textbf{X}$ such that $\textbf{U}_{1} \in \mathbb{R}^{\frac{H}{8} \times \frac{W}{8} \times 2M}$, where $M$ is a variable that controls the length of $\textbf{U}_{1}$.

\begin{figure}[htbp]
\begin{center}
\centerline{\includegraphics[width=0.85\linewidth]{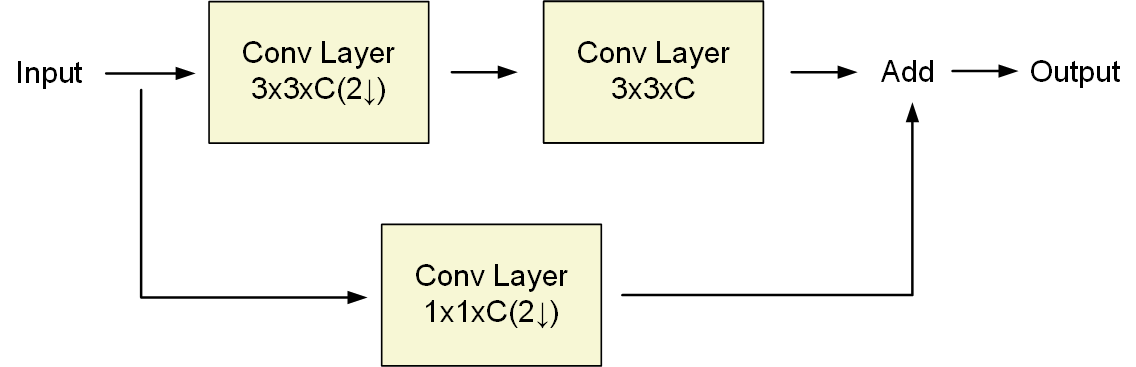}}
\caption{The structure of the residual block with down-sampling.}
\label{fig.resblock}
\end{center}
\vskip -0.3in
\end{figure}

\begin{figure}[htbp]
\begin{center}
\centerline{\includegraphics[width=0.85\linewidth]{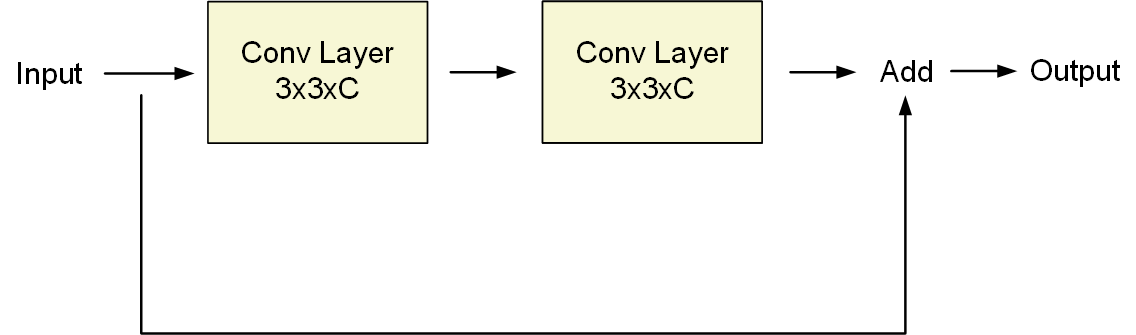}}
\caption{The structure of the residual block without down-sampling.}
\label{fig.resblock2}
\end{center}
\vskip -0.3in
\end{figure}

The structure of the residual block with and without down-sampling is shown in Fig.~\ref{fig.resblock} and Fig.~\ref{fig.resblock2}, respectively.
If the residual block is of the down-sampling type, the input feature maps pass through two convolutional layers and one convolutional layer respectively. 
Otherwise, the separate convolutional layer is replaced by shortcut connection.
The outputs of these two paths are then added together to form the output of a residual block.

\begin{figure}[t]
\begin{center}
\centerline{\includegraphics[width=1\linewidth]{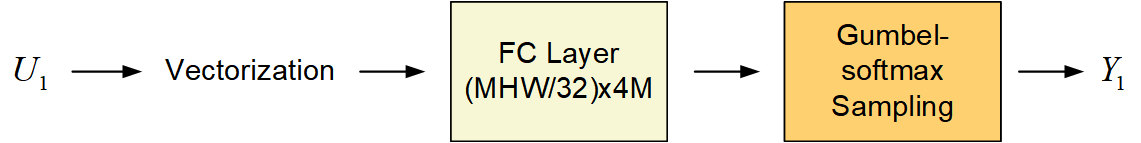}}
\caption{The structure of the 4-QAM modulator.}
\label{fig.3}
\end{center}
\vskip -0.2in
\end{figure}

Since our system is a digital SemCom system, 
directly mapping analog symbols $\textbf{U}_{1}$ onto discrete 4-QAM constellations would result in a non-differentiability issue.
To address this, we adopt the NN-based probabilistic modulator proposed in \cite{bo2024jcm} along with a differentiable sampling technique to process $\textbf{U}_{1}$.
These two parts make up our 4-QAM modulator, and the network structure is shown in Fig.~\ref{fig.3}. 
In the proposed 4-QAM modulator, $\textbf{U}_{1} \in \mathbb{R}^{\frac{H}{8} \times \frac{W}{8} \times 2M} $ is first vectorized to $\bar{\textbf{U}}_{1} \in \mathbb{R}^{\frac{1}{32}MHW \times 1} $. 
Then, an NN-based probabilistic modulator, consisting of one fully connected layer, maps $\bar{\textbf{U}}_{1}$ into a sequence of probability distributions $\left( p^{i1}, p^{i2}, p^{i3}, p^{i4} \right)$ with each element corresponding to the probability of choosing the constellation symbols $\left( 1+1j, 1-1j, -1+1j, -1-1j \right)$, respectively, where $i=1, ..., M$. 
Here $ \frac{1}{32}MHW \times 4M$ below the fully connected layer represents its configuration, where $\frac{1}{32}MHW$ is the number of input neurons, and $4M$ is the number of output neurons.
Then, based on this probability distribution, the constellation symbols are sampled using Gumbel-softmax method \cite{jang2016categorical}, which prevents the non-differentiability problem during sampling.
Following this process, we modulate the semantic information $\textbf{U}_{1}$ into a 4-QAM constellation sequence $\textbf{Y}_{1} \in \mathbb{C}^{M \times 1}$, serving as our outer constellation sequence.

\subsubsection{The Inner Constellation Sequence and Superposition Constellation Map}
The inner constellation sequence, denoted as $\textbf{Y}_{2} \in \mathbb{C}^{M \times 1}$, is generated randomly. For each symbol within the inner constellation sequence, we select a constellation symbol from the set $\left(1+1j, 1-1j, -1+1j, -1-1j\right)$ with equal probability. The length of $\textbf{Y}_{2}$ matches that of $\textbf{Y}_{1}$. 
The design details of the superposition constellation map will be delineated in Section IV.
Specifically, our proposed security approach based on superposition modulation is quite different from the traditional approach \cite{rendon2018nested,wang2013comparison}.
The traditional superposition code scheme employs channel coding, where the information is associated with the codeword randomly and does not require a specific mapping method. 
%
In contrast, our proposed scheme utilizes joint coding-modulation (JCM), which necessitates designing the entire process of associating the source signal with the codeword. 
Moreover, designing a deep learning based superposition code scheme in this context is challenging due to the non-differentiability of digital modulation. In this paper, we address this problem by learning the probability distribution of the outer constellation sequence.

\subsubsection{Semantic Decoder}

\begin{figure}[t]
\begin{center}
\centerline{\includegraphics[width=0.9\linewidth]{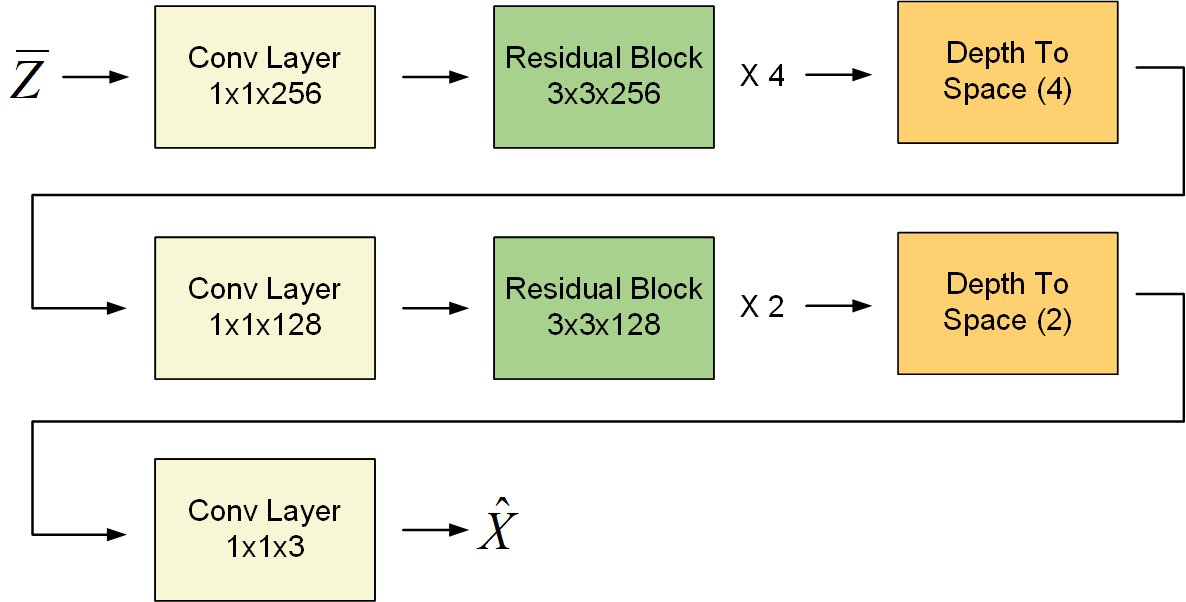}}
\caption{The structure of the semantic decoder.}
\label{fig.4}
\end{center}
\vskip -0.2in
\end{figure}

We assume that the eavesdropper copies the network structure of the semantic decoder of the legitimate user to use as his/her own in order to steal the source image, which is a standard assumption in the study of secure SemCom. Therefore, the structure of the semantic decoder of the eavesdropper is the same with that of the legitimate user. 
The semantic decoder generates the recovered image $\hat{\textbf{X}}$ based on the received semantic information $\bar{\textbf{Z}}$. 
The network structure of the semantic decoder is shown in Fig.~\ref{fig.4},
which consists of three convolutional layers, six residual blocks and two \textit{depth to space} modules. 
The \textit{depth to space} module performs up-sampling, allowing the semantic decoder to recover the source image.
The number shown in \textit{depth to space} box represents its configuration. 
If it is $q$, the number of the feature maps is reduced by a factor of $2^{q}$ and the height and width of the feature maps are increased by a factor of $q$.
The output of the semantic decoder is the recovered source image $\hat{\textbf{X}} \in \mathbb{R}^{H \times W \times 3}$.

\subsection{Training Strategy}
We introduce a two-stage training strategy to train our proposed secure digital SemCom system in this section. 
Unlike other studies, we do not consider preventing the eavesdropper from being able to recover the source data as an optimization goal during the training of the legitimate encoder-decoder pair (the legitimate network), due to the fact that the eavesdropper usually does not cooperate with the legitimate user during the training stage. 
In our setup, the network architecture of the eavesdropper's semantic decoder is identical to that of the legitimate user's semantic decoder.
We also assume that the eavesdropper has access to the entire training dataset while optimizing his/her semantic decoder. However, the eavesdropper does not have access to the pre-trained parameters of the legitimate user's semantic decoder. 
Additionally, the eavesdropper is aware of the PAC value and knows that the received constellation sequence is a superposition of two 4-QAM constellation sequences.
These assumptions are relatively strong for the eavesdropper, and this setup is specifically chosen to effectively highlight the security advantages of our proposed system.

In the first training stage, we train only the legitimate semantic encoder-decoder pair, including the 4-QAM modulator. 
The loss function of the first training stage can be written as 
\begin{equation}
\mathcal{L}_{1}= MSE(\textbf{X},\hat{\textbf{X}}_{1}),
\end{equation}
where $MSE$ is the mean square error (MSE) function which measures the quality of the recovered image. 
Note that we do not consider the eavesdropper's network in this stage.

In the second training stage, we train only the semantic decoder of the eavesdropper. 
We assume that the training data that the eavesdropper has are the noisy transmitted signal $\textbf{Y}_{2}$ and the ground truth $\textbf{X}$. 
The loss function of the second training stage is 
\begin{equation}
\mathcal{L}_2=\ MSE(\textbf{X},\hat{\textbf{X}}_{2}),
\end{equation}
in order for the eavesdropper to optimize his/her semantic encoder.

\section{Constellation Map Design}
As aforementioned, the outer and inner 4-QAM constellation sequences, $\textbf{Y}_{1}$ and $\textbf{Y}_{2}$, are superposed to form a 16-QAM constellation sequence $\textbf{Y} = \sqrt{a} \cdot \textbf{Y}_1 + \sqrt{1 - a} \cdot \textbf{Y}_2$ controlled by a power allocation coefficient (PAC) $a$.
Here, we set the total average transmit power to 1, and $a \in (0,0.5)$. Then $a$ and $1-a$ denote the power allocated to the outer and inner constellation sequences, respectively. 

\begin{figure}[htbp]
\begin{center}
\centerline{\includegraphics[width=0.75\linewidth]{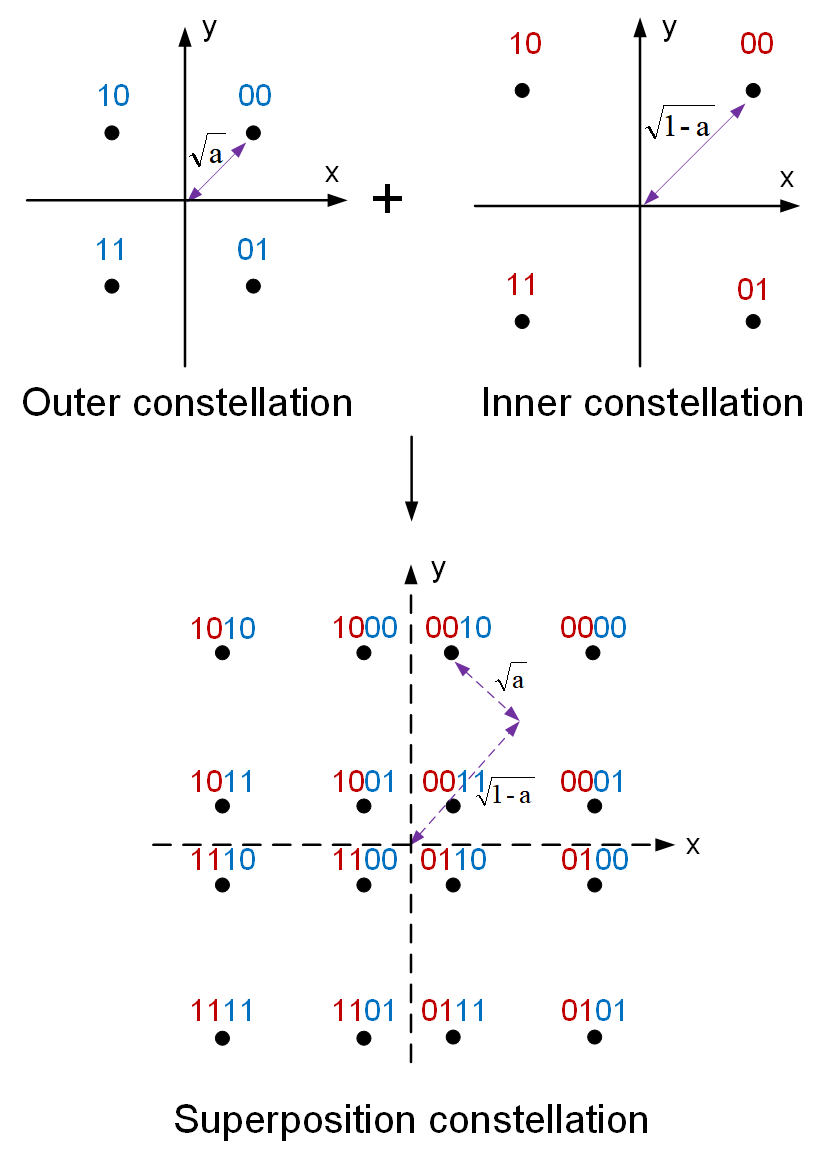}}
\caption{4-QAM + 4-QAM constellation with a power allocation coefficient $a$.}
\label{fig.6}
\end{center}
\vskip -0.3in
\end{figure}

We illustrate in Fig.~\ref{fig.6} the outer and inner 4-QAM constellation maps, as well as the resulted 16-QAM superposition constellation map. 
Since the inner constellation points do not carry valid information, the goal of both the legitimate user and the eavesdropper is to decode the outer constellation symbols.

Different power allocation coefficients result in distinct superposition constellation sequences, leading to varying SEPs when decoding the outer constellation symbols for two users. 
Different channel SNRs of the legitimate user and the eavesdropper also contribute to the difference in the SEP between the two users. Therefore, our objective is to analyze the SEP curves for both the legitimate user and the eavesdropper. 
We aim to identify the optimal $a$ that maximizes the SEP of the eavesdropper while minimizing the SEP of the legitimate user. 
By doing so, we can control the information leakage to the eavesdropper.
In this paper, we assume that both the transmitter and the legitimate user have knowledge of the eavesdropper's channel SNR.

\subsection{SEP Curves with respect to Power Allocation Coefficient}
In this subsection, our goal is to derive the SEP of decoding the outer constellation points as a function of the power allocation coefficient $a$ and the standard deviation $\sigma$ of the AWGN for both users. We calculate the SEP by determining the symbol correctness probability (SCP) of users when decoding the four outer constellation symbols. The maximum likelihood (ML) detector is employed to compute the SCP. It is essential to note that the noise follows a Gaussian distribution with a probability density function of $\mathcal{N}(x;0,\sigma^2)$. Moreover, for any received constellation point, we assume that it is subjected to noise that is independent on the real and imaginary axes.

\begin{figure}[t]
\begin{center}
\centerline{\includegraphics[width=0.75\linewidth]{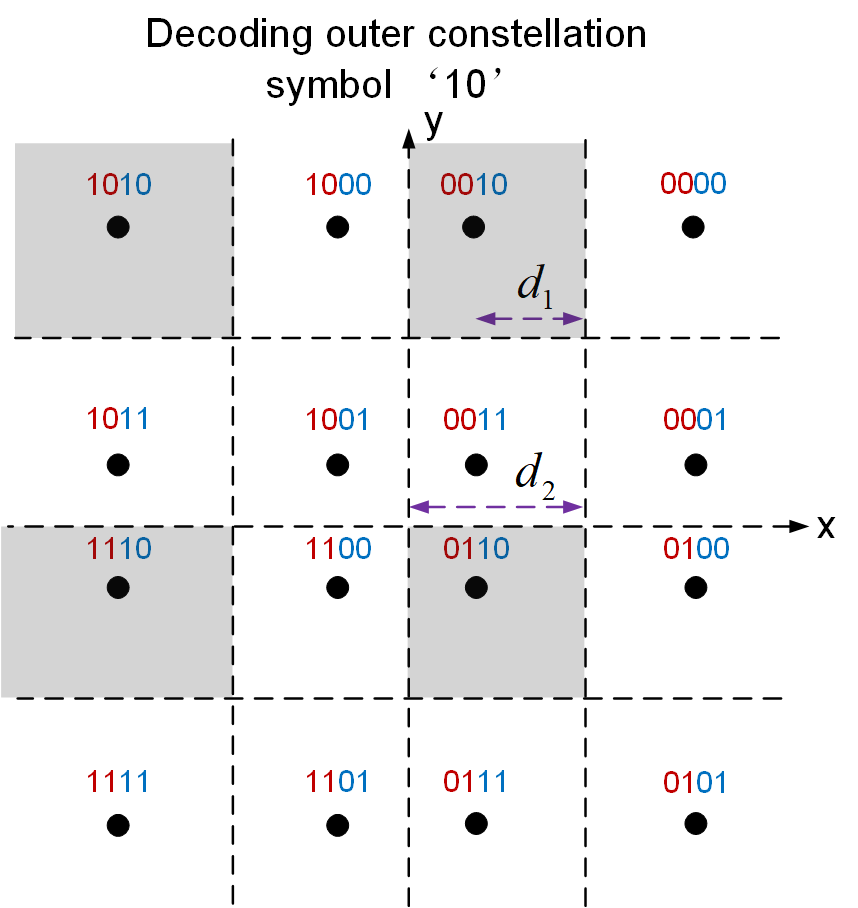}}
\caption{Illustration of decoding the outer constellation symbol `10'.}
\label{fig.7}
\end{center}
\vskip -0.3in
\end{figure}

First, we analyze the case when the sent inner constellation symbol is `00'.
Let us consider the example of decoding the outer constellation symbol `10', as illustrated in Fig.~\ref{fig.7}.
The distance of the inner constellation points from the origin is $\sqrt{1-a}$, and the distance of the outer constellation points from the inner constellation points is $\sqrt{a}$, where $a \in (0,0.5)$.
Thus, we have $d_{1}= \sqrt{a/2}$, $d_{2}= \sqrt{(1-a)/2}$, and $d_{2}>d_{1}$. 


Let $SCP_{s_2}^{s_1}$ denote the probability that the received constellation point falls within the region ${s_2}$, when the sent superposed symbol is $s_1$. For example, $SCP_{0010}^{0010}$ denotes the probability the received symbol falls within the upper right grey region in Fig.~\ref{fig.7} when the sent symbol is $0010$. We note that outer constellation symbol `10' can be successfully decoded as long as the received constellation points are within the four regions of `0010', `1010', `1110', and `0110'. Since the sent constellation symbol is `0010', we take its position as the origin of the maximum likelihood discrimination and calculate $SCP_{0010}^{0010}$, $SCP_{1010}^{0010}$, $SCP_{1110}^{0010}$ and $SCP_{0110}^{0010}$. 

We then have 
\begin{equation}
 SCP_{0010}^{0010} = \int_{-(d_2-d_1)}^{d_1} \int_{-d_1}^{\infty} f_{X,Y}(x, y) \, dy \, dx,   
 \label{eq10}
\end{equation}
where $f_{X,Y}(x, y)$ is the joint probability density function of the Gaussian noise. 
Since the Gaussian noise on the real and imaginary axes in the constellation map is assumed to be independent of each other, $f_{X,Y}(x, y)$ is the product of the two marginal probability density functions, i.e., $f_{X,Y}(x,y) = f_X(x) \cdot f_Y(y) = \mathcal{N}(x;0,\sigma^2) \cdot \mathcal{N}(y;0,\sigma^2)$. 
This yields
%
\begin{equation}
\begin{split}
& SCP_{0010}^{0010} = \int_{-(d_2-d_1)}^{d_1} \mathcal{N}(x; 0, \sigma^2) \, dx \cdot \int_{-d_1}^{\infty} \mathcal{N}(y; 0, \sigma^2) \, dy \\
&= \int_{-(d_2-d_1)}^{d_1} \frac{1}{\sqrt{2\pi}} exp\left(-\frac{\left(\frac{x}{\sigma}\right)^2}{2}\right) d\left(\frac{x}{\sigma}\right) \cdot \\
& \int_{-d_1}^{\infty} \frac{1}{\sqrt{2\pi}} exp\left(-\frac{\left(\frac{y}{\sigma}\right)^2}{2}\right) d\left(\frac{y}{\sigma}\right) \\
&= \left[ \int_{-(d_2-d_1)}^{\infty} \frac{1}{\sqrt{2\pi}} exp\left(-\frac{\left(\frac{x}{\sigma}\right)^2}{2}\right) d\left(\frac{x}{\sigma}\right) + \right. \\
&  \left. \int_{\infty}^{d_1} \frac{1}{\sqrt{2\pi}} exp\left(-\frac{\left(\frac{x}{\sigma}\right)^2}{2}\right) d\left(\frac{x}{\sigma}\right) \right] \\
&\cdot \int_{-d_1}^{\infty} \frac{1}{\sqrt{2\pi}} exp\left(-\frac{\left(\frac{y}{\sigma}\right)^2}{2}\right) d\left(\frac{y}{\sigma}\right) \\
&= \left[ \int_{-(d_2-d_1)}^{\infty} \frac{1}{\sqrt{2\pi}} exp\left(-\frac{\left(\frac{x}{\sigma}\right)^2}{2}\right) d\left(\frac{x}{\sigma}\right) - \right. \\
&  \left. \int_{d_1}^{\infty} \frac{1}{\sqrt{2\pi}} exp\left(-\frac{\left(\frac{x}{\sigma}\right)^2}{2}\right) d\left(\frac{x}{\sigma}\right) \right] \\
&\cdot \int_{-d_1}^{\infty} \frac{1}{\sqrt{2\pi}} exp\left(-\frac{\left(\frac{y}{\sigma}\right)^2}{2}\right) d\left(\frac{y}{\sigma}\right) \\
&= \left[ Q\left(\frac{-(d_2-d_1)}{\sigma}\right) - Q\left(\frac{d_1}{\sigma}\right) \right] \cdot Q\left(\frac{-d_1}{\sigma}\right),
\end{split}
\label{eq16}
\end{equation}
where $Q(x) = \frac{1}{\sqrt{2\pi}} \int_{x}^{\infty} exp\left(-\frac{u^2}{2}\right) du$, and $\sigma$ is the standard deviation of the channel noise. 
In the final step of equation (\ref{eq16}), the variable $u$ in the Q-function corresponds to $\frac{x}{\sigma}$ and $\frac{y}{\sigma}$, respectively. Therefore, in the final expression of the Q-functions, the denominator $\sigma$ consistently appears in the argument, while the numerator of each Q-function is determined by the lower limit of the integrals.
Specifically, for the first Q-function in the last line, the argument is $\frac{-(d_2-d_1)}{\sigma}$; for the second Q-function, it is $\frac{d_1}{\sigma}$; and for the third Q-function, it is $\frac{-d_1}{\sigma}$. 
These arguments correspond to the integration bounds for $\frac{x}{\sigma}$ and $\frac{y}{\sigma}$.

We then calculate the probability of the received symbol in the `1010' region, i.e., the upper left grey region in Fig.~\ref{fig.7}, when the sent symbol is  `0010'. We have
\begin{equation}
\begin{split}
& SCP_{1010}^{0010} = \int_{-\infty}^{-(2d_2-d_1)} \mathcal{N}(x; 0, \sigma^2) \, dx \cdot \int_{-d_1}^{\infty} \mathcal{N}(y; 0, \sigma^2) \, dy \\
& =  Q(\frac{2d_2-d_1}{\sigma}) \cdot Q(\frac{{-d}_1}{\sigma}).
\end{split}
\label{eq17}
\end{equation}

Similarly, the SCP of decoding the constellation symbol `1110' and `0110' can be written as 
\begin{equation}
\begin{split}
& SCP_{1110}^{0010} = \int_{-\infty}^{-(2d_2-d_1)} \mathcal{N}(x; 0, \sigma^2) \, dx \cdot \\ 
&  \int_{-(2d_2+d_1))}^{-(d_1+d_2)} \mathcal{N}(y; 0, \sigma^2) \, dy \\
& =  Q(\frac{2d_2-d_1}{\sigma}) \cdot \left[ Q(\frac{-(2d_2+d_1)}{\sigma}) - Q(\frac{-(d_1+d_2)}{\sigma})\right]. \\
\end{split}
\label{eq18}
\end{equation}

\begin{equation}
\begin{split}
& SCP_{0110}^{0010} = \int_{-(d_2-d_1)}^{d_1} \mathcal{N}(x; 0, \sigma^2) \, dx \cdot \int_{-(2d_2+d_1)}^{-(d_1+d_2)} \mathcal{N}(y; 0, \sigma^2) \, dy \\
& =  \left[Q(\frac{-(d_2-d_1)}{\sigma})-Q(\frac{d_1}{\sigma}) \right] \cdot \\ 
& \left[ Q(\frac{-(2d_2+d_1)}{\sigma}) - Q(\frac{-(d_1+d_2)}{\sigma})\right]. \\
\end{split}
\label{eq19}
\end{equation}

Therefore, according to equations (\ref{eq16})-(\ref{eq19}), the SCP of successfully decoding the outer constellation symbol `10' when the sent symbol is `0010', can be given as 
\begin{equation}
{SCP}_{10}^{0010}={SCP}_{0010}^{0010}+{SCP}_{1010}^{0010}+{SCP}_{1110}^{0010}+{SCP}_{0110}^{0010}.
\label{eq20}
\end{equation}

Following the same process, we give the SCPs of successfully decoding the outer constellation symbols `00', `11', `01', when  the sent symbols are `0000', `0011', `0001', respectively.

\begin{equation}
\begin{split}
{SCP}_{00}^{0000}={SCP}_{0000}^{0000}+{SCP}_{1000}^{0000}+{SCP}_{1100}^{0000}+{SCP}_{0100}^{0000},
\end{split}
\label{eq21}
\end{equation}
where
\begin{equation}
\begin{split}
& SCP_{0000}^{0000} = { Q(\frac{-d_1}{\sigma})}^{2}. \\
& SCP_{1000}^{0000} = \left[ Q(\frac{-(2d_2+d_1)}{\sigma}) - Q(\frac{-(d_1+d_2)}{\sigma})\right] \cdot Q(\frac{-d_1}{\sigma}). \\
& SCP_{1100}^{0000} = {\left[ Q(\frac{-(2d_2+d_1)}{\sigma}) - Q(\frac{-(d_1+d_2)}{\sigma})\right]}^{2} . \\
\end{split}
\end{equation}
\begin{equation}
\begin{split}
& SCP_{0100}^{0000} = \left[ Q(\frac{-(2d_2+d_1)}{\sigma}) - Q(\frac{-(d_1+d_2)}{\sigma})\right] \cdot Q(\frac{-d_1}{\sigma}). 
\end{split}
\tag{\theequation}
\end{equation}

\begin{equation}
\begin{split}
{SCP}_{11}^{0011}={SCP}_{0011}^{0011}+{SCP}_{1011}^{0011}+{SCP}_{1111}^{0011}+{SCP}_{0111}^{0011},
\end{split}
\label{eq23}
\end{equation}
where
\begin{equation}
\begin{split}
& SCP_{0011}^{0011} = {\left[ Q(\frac{-(d_2-d_1)}{\sigma}) - Q(\frac{d_1}{\sigma})\right]}^{2}. \\
& SCP_{1011}^{0011} = Q(\frac{2d_2-d_1}{\sigma}) \cdot \left[Q(\frac{-(d_2-d_1)} {\sigma})-Q(\frac{d_1} {\sigma})\right] . \\
& SCP_{1111}^{0011} = { Q(\frac{2d_2-d_1}{\sigma})}^{2} . \\
& SCP_{0111}^{0011} = Q(\frac{2d_2-d_1}{\sigma}) \cdot \left[Q(\frac{-(d_2-d_1)} {\sigma})-Q(\frac{d_1} {\sigma})\right] . \\
\end{split}
\end{equation}

\begin{equation}
\begin{split}
{SCP}_{01}^{0001}={SCP}_{0001}^{0001}+{SCP}_{1001}^{0001}+{SCP}_{1101}^{0001}+{SCP}_{0101}^{0001},
\end{split}
\label{eq25}
\end{equation}
where
\begin{equation}
\begin{split}
& SCP_{0001}^{0001} = Q(\frac{-d_1}{\sigma}) \cdot \left[ Q(\frac{-(d_2-d_1)}{\sigma}) - Q(\frac{d_1}{\sigma})\right]. \\
& SCP_{1001}^{0001} = \left[ Q(\frac{-(2d_2+d_1)}{\sigma}) - Q(\frac{-(d_1+d_2)}{\sigma})\right] \cdot \\
& \left[Q(\frac{-(d_2-d_1)}{\sigma})-Q(\frac{d_1}{\sigma}) \right] . \\
& SCP_{1101}^{0001} = \left[ Q(\frac{-(2d_2+d_1)}{\sigma}) - Q(\frac{-(d_1+d_2)}{\sigma})\right] \cdot \\
& Q(\frac{2d_2-d_1}{\sigma}) . \\
& SCP_{0101}^{0001} = Q(\frac{-d_1}{\sigma}) \cdot Q(\frac{2d_2-d_1}{\sigma}). \\
\end{split}
\end{equation}

Extensive simulation results reveal that the probabilities of sending the four outer constellation points are approximately equal.
Hence, when the sent inner constellation symbol is `00', according to equations (\ref{eq20}), (\ref{eq21}), (\ref{eq23}), and (\ref{eq25}), the SEP of decoding the sent outer constellation symbols is 
\begin{equation}
SEP^{00}=1-\frac{({SCP}_{10}^{0010}+{SCP}_{00}^{0000}+{SCP}_{11}^{0011}+{SCP}_{01}^{0001})}{4}. 
\label{eq27}
\end{equation}

Subsequently, we compute the SEP of decoding the outer constellation symbols when the sent inner constellation symbols are `01', `10' and `11', respectively.
Note that for different inner constellation symbols, the SCP of decoding the outer constellation symbols has certain symmetry. For example, the SCP of decoding the outer constellation symbol `00' when the sent inner constellation symbol is `01' is the same as the SCP of decoding the outer constellation symbol `10' when the sent inner constellation symbol is `00', i.e., ${SCP}_{00}^{0100}={SCP}_{10}^{0010}$.
This is equivalent to rotating the constellation map 90 degrees clockwise.
It easily follows that the SEP of decoding the outer constellation symbols is the same across all four inner constellation symbols, which can be written as 
\begin{equation}
SEP^{00}=SEP^{01}=SEP^{10}=SEP^{11}.
\label{eqend}
\end{equation}
Then we obtain the SEP of decoding the outer constellation symbols as shown in \textit{Theorem~\ref{theo1}}.

\begin{theorem}

According to equations (\ref{eq10})-(\ref{eqend}), when the superposition constellation sequence is 16-QAM constellation sequence, the formula for SEP of decoding the outer 4-QAM constellation symbols can be written as 
\begin{equation}
SEP=\frac{(SEP^{00}+SEP^{01}+SEP^{10}+SEP^{11})}{4}=SEP^{00}, 
\label{SEP-calculate}
\end{equation}
\label{theo1}
where $SEP^{00}$ is given in equation (\ref{eq27}).

\end{theorem}

\begin{figure}[t]
\begin{center}
\centerline{\includegraphics[width=1\linewidth]{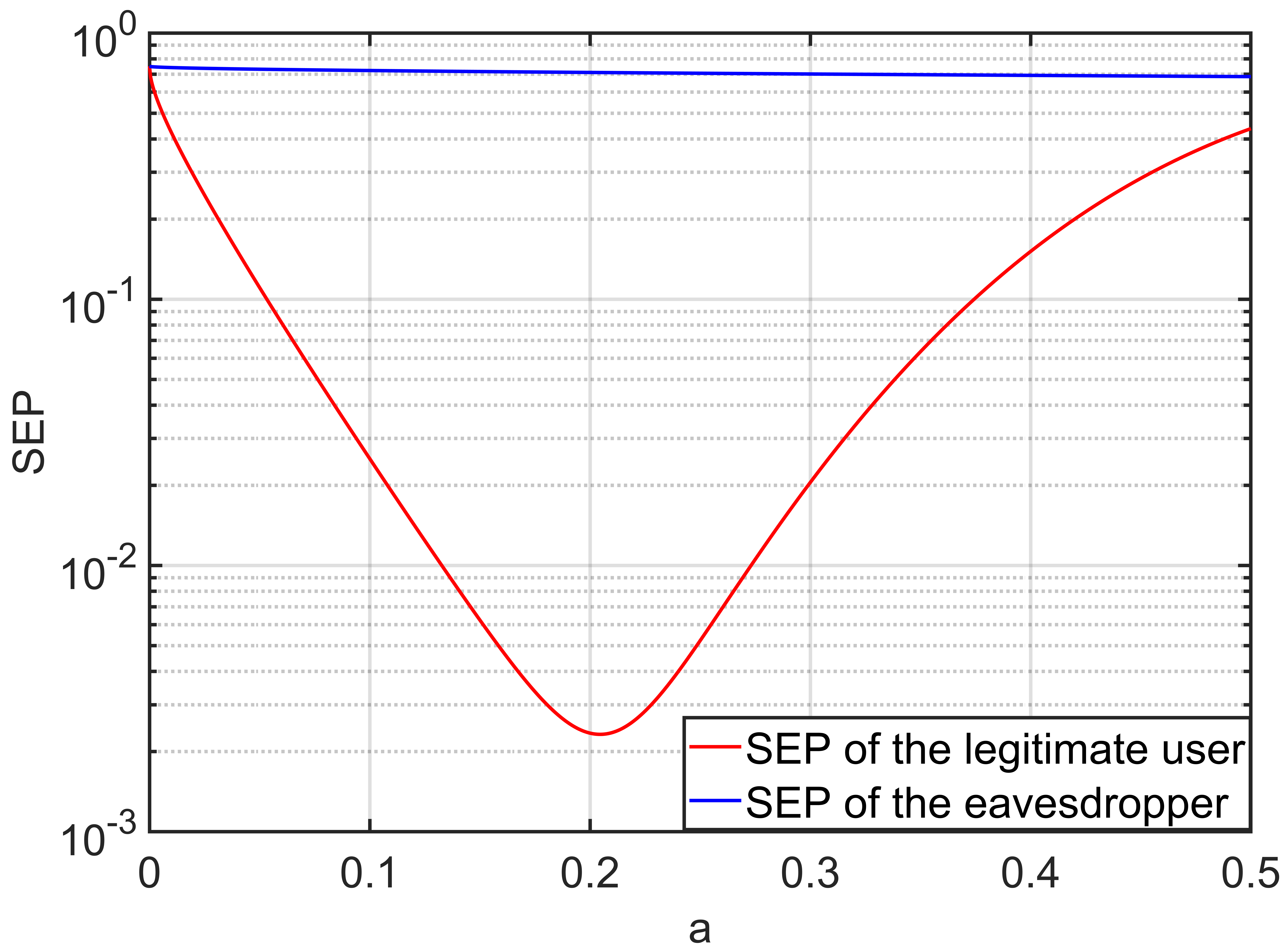}}
\caption{The SEP of the legitimate user and the eavesdropper when $\mathrm{SNR}_{\mathrm{leg}}=20\mathrm{dB}$ and $\mathrm{SNR}_{\mathrm{eve}}=-10\mathrm{dB}$.}
\label{fig.8}
\end{center}
\vskip -0.3in
\end{figure}

Recall that the channel SNR of the legitimate user and the eavesdropper is denoted by $\mathrm{SNR}_{\mathrm{leg}}$ and $\mathrm{SNR}_{\mathrm{eve}}$, respectively. We plot the curves of the SEP by the legitimate user and the eavesdropper with regards to the power allocation coefficient $a$ in Fig.~\ref{fig.8}, when $\mathrm{SNR}_{\mathrm{leg}}=20\mathrm{dB}$ and $\mathrm{SNR}_{\mathrm{eve}}=-10\mathrm{dB}$.
We can observe that the SEP of the eavesdropper remains high and decreases very slowly with $a$ since the eavesdropper suffers from poor channel. It can also be observed that the SEP of the legitimate user initially decreases rapidly and then increases rapidly. Based on this finding, we select a power allocation coefficient $a$ such that the SEP of the legitimate user is relatively low, while the SEP of the eavesdropper is high enough. 
This can nearly prevent information leakage to the eavesdropper and makes it less likely for them to recover useful information.
For example, in Fig.~\ref{fig.8}, when we choose $a=0.055$, the SEP of the eavesdropper is 73\%, making it almost impossible to recover any valid information. 
This is because in the case of 4-QAM, which has four symbols, an error rate of 75\% represents the worst-case scenario, equivalent to random guessing. A 73\% SEP is very close to this 75\%, meaning the eavesdropper is essentially guessing the symbols at random, making it nearly impossible to recover any meaningful information.
Meanwhile, the SEP of the legitimate user is only 9.4\%, which, relative to the eavesdropper, is significantly lower, ensuring that the legitimate user's performance is much better.
Based on the above SEP results, to find the optimal power allocation coefficient $a$ under a certain legitimate user-eavesdropper SNR pair, we formulate an optimization algorithm as follows
\begin{equation}\label{optimization}
\min \ {SEP}_{\rm leg}\ s.t.\ {SEP}_{\rm eve} \geq b,
\end{equation}
where ${SEP}_{\rm leg}$ and ${SEP}_{\rm eve}$ are the SEP of the legitimate user and the eavesdropper, respectively, and $b$ is the minimum value of ${SEP}_{\rm eve}$, limiting information leakage to the eavesdropper.
%
$b$ is referred to as the minimum SEP (MSEP), also known as the threshold. 
A higher value of MSEP indicates a more secure system with better data privacy for the transmitter, but it comes at the cost of performance loss for the legitimate user. 
By adjusting $b$, we can control the extent of information leakage to the eavesdropper, achieving security that is both controllable and quantifiable.
%
The hyperparameters PAC and MSEP are only related to the encoding, modulation and superposition of semantic information at the transmitter. 
Meanwhile, the eavesdropper possesses only the semantic decoder, thus the eavesdropper's network structure is independent of PAC and MSEP.
Based on the proposed optimization algorithm, we give the optimal value of PAC under different legitimate user-eavesdropper SNR pairs and different values of MSEP, as shown in Table~\ref{tab1}. Note that in this paper, we assume the channel SNR assigned to each user represents the optimal channel SNR achievable by the legitimate user and the eavesdropper.
Furthermore, since the eavesdropper and the legitimate user employ the same decoding method and the semantic information is not encrypted, our system, consistent with most existing studies, relies on the assumption of a difference in channel SNR between the legitimate user and the eavesdropper.
\begin{table}[h]
\caption{The optimal value of PAC under different legitimate user-eavesdropper SNR pairs and different values of MSEP.}
\centering
\begin{tabular}{c|c|c|c}
\hline
$\mathrm{SNR}_{\mathrm{leg}}$ (dB) & $\mathrm{SNR}_{\mathrm{eve}}$ (dB) & MSEP $b$ (\%) & PAC $a$ \\ \hline
\multirow{20}{*}{20}                          & \multirow{4}{*}{-15}                            & 74             & 0.040      \\ \cline{3-4}
                            &                                & 73             & 0.158      \\ \cline{3-4}
                            &                                & 72             & 0.347      \\ \cline{3-4}
                            &                                & 71             & -      \\ \cline{2-4}
                         &       \multirow{4}{*}{-10}                         & 74             & 0.014      \\ \cline{3-4}
                            &                                & 73             & 0.055      \\ \cline{3-4}
                            &                                & 72             & 0.121      \\ \cline{3-4}
                            &                                & 71             & 0.209      \\ \cline{2-4}
                         &        \multirow{4}{*}{-5}                         & 74             & 0.006      \\ \cline{3-4}
                            &                                & 73             & 0.025      \\ \cline{3-4}
                            &                                & 72             & 0.051      \\ \cline{3-4}
                            &                                & 71             & 0.087      \\ \cline{2-4}
                           &     \multirow{4}{*}{0}                           & 74             & 0.003      \\ \cline{3-4}
                            &                                & 73             & 0.014      \\ \cline{3-4}
                            &                                & 72             & 0.030      \\ \cline{3-4}
                            &                                & 71             & 0.052      \\ \cline{2-4}
                        &      \multirow{4}{*}{5}                           & 74             & 0.001      \\ \cline{3-4}
                            &                                & 73             & 0.004      \\ \cline{3-4}
                            &                                & 72             & 0.009      \\ \cline{3-4}
                            &                                & 71             & 0.017      \\ \cline{2-4} \hline
\end{tabular}
\label{tab1}
\vskip -0.1in
\end{table}

To clarify the SEP calculation process, we further present the intermediate steps involved in the calculation of SEP. 
We select a scenario from Table~\ref{tab1} and clearly displayed the exact values of each component of the SEP in Table~\ref{SEP_cal_1}. 
The chosen scenario is $\mathrm{SNR}_{\mathrm{leg}}=20\mathrm{dB}$, $\mathrm{SNR}_{\mathrm{eve}}=-15\mathrm{dB}$, and PAC = 0.040. 
According to \textit{Theorem~\ref{theo1}} and equation (\ref{SEP-calculate}), the SEP is equivalent to $SEP^{00}$. 
Therefore, we calculate the SEP based on equation (\ref{eq27}), where the four components in equation (\ref{eq27}) are derived from equations (\ref{eq20}), (\ref{eq21}), (\ref{eq23}), and (\ref{eq25}), respectively.

\begin{table}[ht]
\caption{Intermediate values for SEP calculation when $\mathrm{SNR}_{\mathrm{leg}}=20\mathrm{dB}$, $\mathrm{SNR}_{\mathrm{eve}}=-15\mathrm{dB}$, and PAC = 0.040.}
\centering
\begin{minipage}{0.48\textwidth}
\centering
\caption*{(a) SEP calculation for the eavesdropper}
\begin{tabular}{c|c|c|c|c}
\hline
Variable & \textbf{\( SCP_{0010}^{0010} \)} & \textbf{\( SCP_{0000}^{0000} \)} & \textbf{\( SCP_{0011}^{0011} \)} & \textbf{\( SCP_{0001}^{0001} \)} \\ \hline
Value & 0.0250 & 0.2601 & 0.0024 & 0.0250 \\ \hline
Variable & \textbf{\( SCP_{1010}^{0010} \)} & \textbf{\( SCP_{1000}^{0000} \)} & \textbf{\( SCP_{1011}^{0011} \)} & \textbf{\( SCP_{1001}^{0001} \)} \\ \hline
Value & 0.2104 & 0.0245 & 0.0202 & 0.0024 \\ \hline
Variable & \textbf{\( SCP_{1110}^{0010} \)} & \textbf{\( SCP_{1100}^{0000} \)} & \textbf{\( SCP_{1111}^{0011} \)} & \textbf{\( SCP_{1101}^{0001} \)} \\ \hline
Value & 0.0198 & 0.0023 & 0.1701 & 0.0198 \\ \hline
Variable & \textbf{\( SCP_{0110}^{0010} \)} & \textbf{\( SCP_{0100}^{0000} \)} & \textbf{\( SCP_{0111}^{0011} \)} & \textbf{\( SCP_{0101}^{0001} \)} \\ \hline
Value & 0.0024 & 0.0245 & 0.0202 & 0.2104 \\ \hline
Variable & \textbf{\( SCP_{10}^{0010} \)} & \textbf{\( SCP_{00}^{0000} \)} & \textbf{\( SCP_{11}^{0011} \)} & \textbf{\( SCP_{01}^{0001} \)} \\ \hline
Value & 0.2576 & 0.3115 & 0.2130 & 0.2576 \\ \hline
\multicolumn{5}{c}{SEP Result: 0.7401 } \\ \hline
\end{tabular}
\label{SEP-values1}
\end{minipage}
\hfill
\begin{minipage}{0.48\textwidth}
\centering
\caption*{(b) SEP calculation for the legitimate user}
\begin{tabular}{c|c|c|c|c}
\hline
Variable & \textbf{\( SCP_{0010}^{0010} \)} & \textbf{\( SCP_{0000}^{0000} \)} & \textbf{\( SCP_{0011}^{0011} \)} & \textbf{\( SCP_{0001}^{0001} \)} \\ \hline
Value & 0.8489 & 0.8489 & 0.8489 & 0.8489 \\ \hline
Variable & \textbf{\( SCP_{1010}^{0010} \)} & \textbf{\( SCP_{1000}^{0000} \)} & \textbf{\( SCP_{1011}^{0011} \)} & \textbf{\( SCP_{1001}^{0001} \)} \\ \hline
Value & 0 & 0 & 0 & 0 \\ \hline
Variable & \textbf{\( SCP_{1110}^{0010} \)} & \textbf{\( SCP_{1100}^{0000} \)} & \textbf{\( SCP_{1111}^{0011} \)} & \textbf{\( SCP_{1101}^{0001} \)} \\ \hline
Value & 0 & 0 & 0 & 0 \\ \hline
Variable & \textbf{\( SCP_{0110}^{0010} \)} & \textbf{\( SCP_{0100}^{0000} \)} & \textbf{\( SCP_{0111}^{0011} \)} & \textbf{\( SCP_{0101}^{0001} \)} \\ \hline
Value & 0 & 0 & 0 & 0 \\ \hline
Variable & \textbf{\( SCP_{10}^{0010} \)} & \textbf{\( SCP_{00}^{0000} \)} & \textbf{\( SCP_{11}^{0011} \)} & \textbf{\( SCP_{01}^{0001} \)} \\ \hline
Value & 0.8489 & 0.8489 & 0.8489 & 0.8489 \\ \hline
\multicolumn{5}{c}{SEP Result: 0.1511 } \\ \hline
\end{tabular}
\label{SEP-values2}
\end{minipage}
\label{SEP_cal_1}
\end{table}

\subsection{Equivalent Channel Capacity}

In this subsection, we analyze the gap between the channel capacity of the proposed superposition code based transmission scheme and the wiretap channel capacity \cite{leung1978gaussian}. 
%
This gap represents the error-free transmission rate lost by the legitimate user due to the proposed security scheme under a specified MSEP.
%
Note that the channel capacity is determined by channel SNR according to the Shannon equation. 
Hence, equivalently, we compare the actual equivalent channel SNR based on different MSEPs with the wiretap channel capacity equivalent SNR to evaluate the channel capacity gap. 
We denote the wiretap channel capacity equivalent SNR as $\mathrm{SNR}_{\mathrm{equ}}^{1}$. 
We can calculate the wiretap channel capacity  $C_{\mathrm{wiretap}}$ \cite{wyner1975wire} by 
\begin{equation}
C_{\mathrm{wiretap}}=max\{B\log_2(1+{\rm SNR}_{\rm leg})-B\log_2(1+{\rm SNR}_{\rm eve}),0\},
\end{equation}
where $B$ is the channel bandwidth. We recall that ${\rm SNR}_{\rm leg}$ and ${\rm SNR}_{\rm eve}$ is the channel SNR of the legitimate user and the eavesdropper, respectively. We then have 
\begin{equation}
{B \rm log}_2(1+{\rm SNR}_{\rm equ}^1)={B \rm log}_2(1+{\rm SNR}_{\rm leg})-{B \rm log}_2(1+{\rm SNR}_{\rm eve}), 
\end{equation}
which yields
\begin{equation}
\begin{split}
& {\rm SNR}_{\rm equ}^1=2^{\left(\log_2\left(1+{\rm SNR}_{\rm leg}\right)-\log_2\left(1+{\rm SNR}_{\rm eve}\right)\right)}-1 \\
& =10\log_{10}(2^{\left(\log_2\left(1+{\rm SNR}_{\rm leg}\right)-\log_2\left(1+{\rm SNR}_{\rm eve}\right)\right)}-1)\ (\mathrm{dB}).
\end{split}
\end{equation}


Different MSEPs correspond to different optimal PACs $a=g(b)$ by solving the optimization problem in equation (\ref{optimization}). We denote the actual equivalent channel SNR for a given MSEP by $\mathrm{SNR}_{\mathrm{equ}}^{2}(b)$, which can be calculated by 
\begin{equation}
   \mathrm{SNR}_{\mathrm{equ}}^{2}(b) = 10 \log_{10}\left(\frac{g(b)P}{\sigma_{1}^{2}}\right) \, (\mathrm{dB}).
\end{equation}

\begin{figure}[t]
\begin{center}
\centerline{\includegraphics[width=1\linewidth]{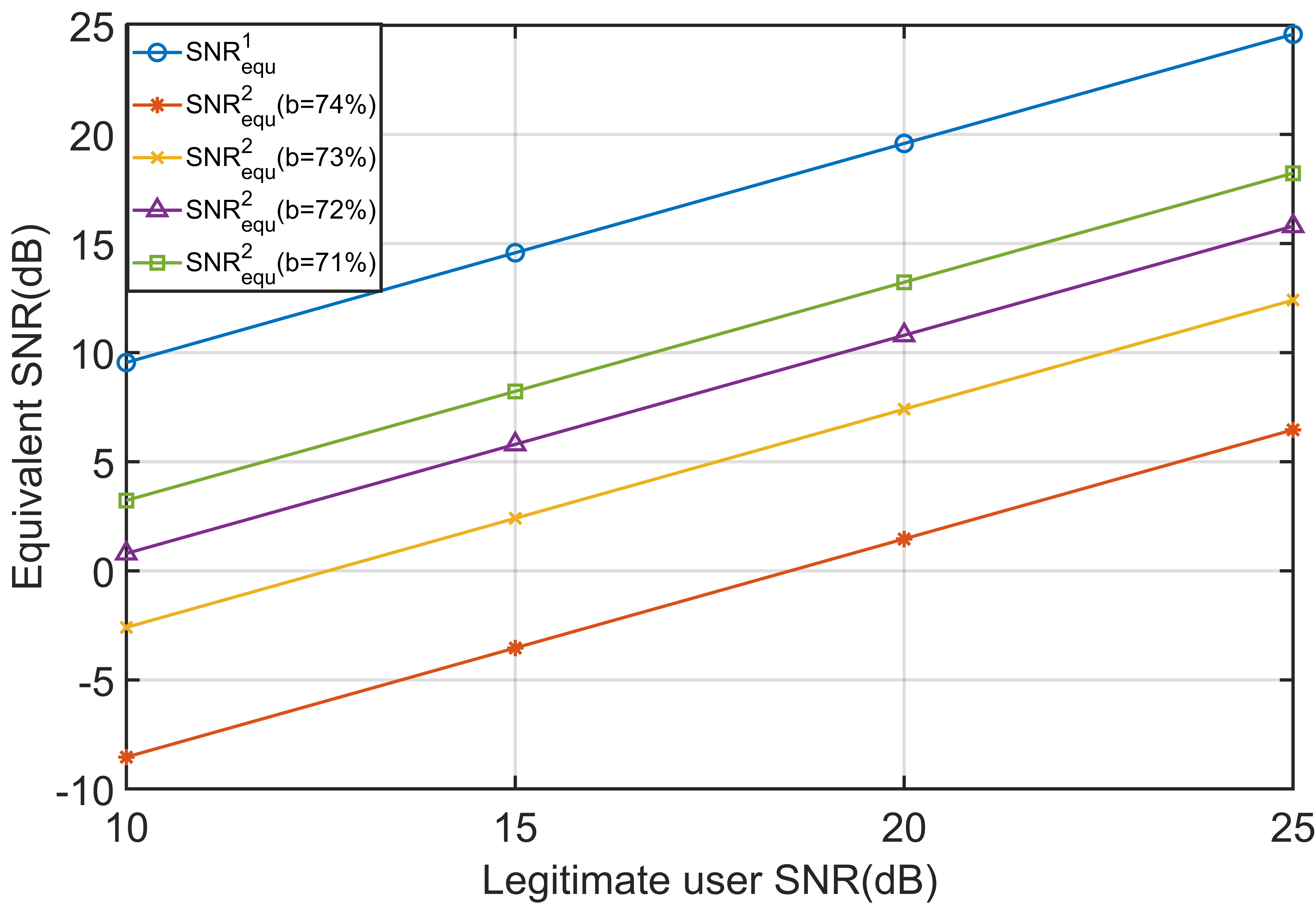}}
\caption{Equivalent channel SNR curves under different $\mathrm{SNR}_{\mathrm{leg}}$ values when $\mathrm{SNR}_{\mathrm{eve}}=-10\mathrm{dB}$.}
\label{fig.9}
\end{center}
\vskip -0.3in
\end{figure}

We assess the capacity loss for security by comparing the actual equivalent channel SNR $\mathrm{SNR}_{\mathrm{equ}}^{2}(b)$ with the wiretap channel capacity equivalent SNR $\mathrm{SNR}_{\mathrm{equ}}^{1}$, as illustrated in Fig.~\ref{fig.9}.
We can observe that the gap between the actual equivalent channel SNR and the wiretap channel capacity equivalent SNR becomes smaller as MSEP decreases. It is important to note that the wiretap channel capacity represents a scenario without security considerations. As MSEP increases, our system sacrifices more channel capacity for enhanced security. Conversely, with a lower MSEP, the channel capacity of our system increases, but the security level decreases. This highlights the inherent trade-off between the security and channel capacity of our system.

\section{Simulation Results}
\subsection{Experimental Settings}
\subsubsection{Dataset}
Our experiments utilize the CIFAR-10 dataset \cite{krizhevsky2009learning}, 
comprising 50,000 training images and 10,000 testing images.
All the images in the dataset are $32 \times 32$ RGB images.

\subsubsection{Compression Ratio}
We denote the number of symbols in the source image by $n$ and the number of \textit{real-valued} symbols transmitted over the channel by $k$. The compression ratio is denoted as $CR=k/2n$. In this paper, the compression ratio falls within the set $CR\in\{1/24,2/24,3/24,4/24,5/24,6/24,12/24\}$.

\subsubsection{Effect of Modulation Order Setting}
For experimental convenience, we have set the outer modulation order to 4-QAM in our proposed system. 
However, it is important to note that this method can be easily extended to higher-order modulation schemes.
Specifically, employing higher-order modulation schemes only increases the number of output neurons in the FC layer of the QAM modulator and the number of input channels in the first convolutional layer of the semantic decoder. Therefore, higher-order modulation does not significantly increase the system's training and inference time, as it only affects these two layers in our proposed architecture.
The primary consequence is a moderate increase in the training time per epoch for end-to-end training as the modulation order increases. 
For instance, as demonstrated in our previous work\cite{bo2024jcm}, increasing the modulation order from 16-QAM to 64-QAM results in only about a 14\% increase in the training time per epoch, while the total number of epochs required for convergence remains unchanged.

On the other hand, the power allocation scheme experiences a slight increase in complexity when higher-order modulation is used.
This is due to the more complex SEP derivation required for both the legitimate user and the eavesdropper when decoding the outer constellation symbols. 
While this may slightly increase the complexity during the system's initial design phase, it does not impact the training or inference process.
In conclusion, higher-order modulation schemes introduce only minimal additional complexity to both the system architecture and training.
While there may be some increase in the complexity of SEP derivation, this is manageable and independent of the training and inference stages.
Therefore, our system can be easily extended to accommodate higher-order modulation schemes.

\subsubsection{Training Settings}

Our experiments are conducted on a single NVIDIA RTX A6000 GPU.
The batch size is set to 256, and during training, we use the Adam optimizer. 
Throughout the training process, the PAC is maintained as a fixed value.
During the initial training stage, we train the legitimate semantic encoder-decoder pair. The learning rate schedule is as follows: $2 \times 10^{-4}$ for 30 epochs, $1 \times 10^{-4}$ for 40 epochs, $5 \times 10^{-5}$ for 40 epochs, and $1 \times 10^{-5}$ for the remaining 40 epochs. 
It's noteworthy that the learning rate of the 4-QAM modulator is half of the specified learning rate. 
In the subsequent training stage, only the semantic decoder of the eavesdropper is trained, with the learning rate following this sequence: $2 \times 10^{-4}$ for 50 epochs, $5 \times 10^{-5}$ for 50 epochs, and $1 \times 10^{-5}$ for the final 50 epochs.

\subsubsection{The Benchmarks}
We utilize four benchmarks for comparison purposes. 
In the first benchmark, Alice transmits a 16-QAM constellation sequence $\textbf{Y}_1$ directly to the legitimate user without employing a superposition code. In this scenario, there is no randomly generated constellation sequence $\textbf{Y}_2$, and no power allocation coefficient $a$ is needed. This benchmark essentially represents a digital SemCom system without specific security design considerations. All other parameters and settings align with the proposed system. 
The performance of this benchmark serves as a experimental lower bound of security under the settings described in this paper.
This benchmark highlights the importance of the inner constellation sequence and the appropriate PAC in enhancing security.
We refer to this benchmark as \textit{16-QAM without Superposition}.

In the second benchmark, we combine the \textit{16-QAM without Superposition} approach with the advanced encryption standard (AES) encryption method. 
In this setup, AES is configured with a 128-bit key, adhering to the AES-128 standard.
The encryption process uses the cipher block chaining (CBC) mode.
Alice and Bob exchange an encryption key in advance, with Alice encrypting the transmitted semantic information using AES before sending it. 
Upon receiving the encrypted data, Bob decrypts it using the shared key. In contrast, Eve, who lacks the encryption key, is unable to decrypt the transmitted semantic information. 
This benchmark offers the advantage that the legitimate user has access to the encryption key, providing a significant security advantage.
However, it also introduces a potential vulnerability, as the system's security could be compromised by quantum attacks.
We refer to this benchmark as \textit{16-QAM with AES Encryption}.

In the third benchmark, we implement the adversarial training method proposed in \cite{marchioro2020adversarial} and \cite{luo2023encrypted}, which is one of the current state-of-the-art solutions in secure SemCom over wiretap channels.
Following the network structure outlined in \cite{marchioro2020adversarial}, we consider an analog SemCom system without modulation and superposition code. The semantic encoder generates the channel input symbols, which are directly fed into the AWGN channel. 
%
In this benchmark, the structure of the legitimate semantic encoder-decoder pair is replaced with the network structure proposed in \cite{bourtsoulatze2019deep}.
This modification is made because \cite{marchioro2020adversarial} and \cite{luo2023encrypted} focus on classification tasks, while our work targets image reconstruction. 
To align with our task while remaining close to their network design, we adopt the widely used SemCom architecture \cite{bourtsoulatze2019deep} for this benchmark.
In addition, the original cross-entropy loss used for classification is replaced with MSE loss to ensure consistency with the training of our proposed system.
%
For this benchmark, we adopt a two-stage training strategy. In the first training stage, an adversarial training strategy is employed, involving a minimax game between the Alice and Bob pair and Eve. Both the legitimate semantic encoder-decoder pair and the semantic decoder of the eavesdropper are trained. The loss function of the first stage is expressed as:
\begin{equation}
\mathcal{L}_{\rm benchmark}^{1}= MSE(\textbf{X},\hat{\textbf{X}}_{1})-\lambda \cdot MSE(\textbf{X},\hat{\textbf{X}}_{2}),
\end{equation}
where $\lambda = 2 \times 10^{-2}$ is a trade-off hyperparameter.
The legitimate network aims to minimize $\mathcal{L}_{\rm benchmark}^{1}$, while the eavesdropper's network (the semantic decoder of the eavesdropper) aims to maximize it. The training follows the approach of generative adversarial networks (GANs), with iterative parameter updates for both networks. 
%
Subsequently, in the second training stage, only the eavesdropper's network is trained to enhance its performance. The loss function of the second stage is given by:
\begin{equation}
\mathcal{L}_{\rm benchmark}^{2}= MSE(\textbf{X},\hat{\textbf{X}}_{2}).
\end{equation} 
All other settings remain consistent with the proposed system, and we refer to this benchmark as \textit{Adversarial Training}.

In the fourth benchmark, both the inner and outer 4-QAM constellation sequences $\textbf{Y}_1$ and $\textbf{Y}_2$ are randomly generated. They are subsequently superposed to generate 16-QAM superposition constellation sequence. This benchmark is used to evaluate the PSNR performance when the signal received by the eavesdropper is a random constellation sequence, aiming to analyze whether our proposed system can nearly achieve the experimental upper bound of security in digital SemCom systems.
The other settings remain consistent with the proposed system. 
We refer to this benchmark as \textit{Random Constellation Sequence}.

\subsection{The Training and Inference Times of the Proposed System}

In this subsection, we report the average training time per image for each epoch across different training stages, conducted on a single NVIDIA RTX A6000 GPU. Additionally, we present the average inference time per image during the inference stage, measured on the same GPU device.
We also calculate the number of image frames that our proposed system can process per second during the inference stage, demonstrating the real-time communication advantages of our proposed system, as shown in Table~\ref{tab2-24} and Table~\ref{tab4-24}. 
Table~\ref{tab2-24} and Table~\ref{tab4-24} display the results for the training and inference times of our proposed system at compression ratios of 2/24 and 4/24, respectively.
The training dataset consists of 50,000 images, while the testing dataset contains 10,000 images.
It is important to note that during the first training stage, the eavesdropper's semantic decoder is frozen. During the second training stage, both the semantic encoder and decoder of the legitimate user are frozen. During the inference stage, the entire network is active.

\begin{table}[ht]
\caption{The training and inference times of our proposed system on a single NVIDIA RTX A6000 GPU at CR = $2/24$.}
\centering
\begin{tabular}{c|c|c}
\hline
Stage & \begin{tabular}[c]{@{}c@{}}Time per\\ image (ms)\end{tabular}& \begin{tabular}[c]{@{}c@{}}Total time\\ per epoch (s)\end{tabular} \\ \hline
Training (Stage 1)  & 0.11 ms  & 5.26 s  \\ \hline
Training (Stage 2)  & 0.08 ms  & 4.00 s  \\ \hline
Inference           & 0.10 ms  & 1.01 s  \\ \hline
\multicolumn{3}{c}{Frames processed per second: 10,000 frames} \\ \hline
\end{tabular}
\label{tab2-24}
\end{table}

\begin{table}[ht]
\caption{The training and inference times of our proposed system on a single NVIDIA RTX A6000 GPU at CR = $4/24$.}
\centering
\begin{tabular}{c|c|c}
\hline
Stage & \begin{tabular}[c]{@{}c@{}}Time per\\ image (ms)\end{tabular}& \begin{tabular}[c]{@{}c@{}}Total time\\ per epoch (s)\end{tabular} \\ \hline
Training (Stage 1)  & 0.13 ms  & 6.40 s  \\ \hline
Training (Stage 2)  & 0.10 ms  & 5.11 s  \\ \hline
Inference           & 0.13 ms  & 1.26 s  \\ \hline
\multicolumn{3}{c}{Frames processed per second: 7,692 frames} \\ \hline
\end{tabular}
\label{tab4-24}
\end{table}

From Table~\ref{tab2-24} and Table~\ref{tab4-24}, it is clear that the processing time per image during both the training and inference stages is exceptionally low.
For the system with CR = $2/24$, the processing time per image in the first and second training stages is just 0.11 ms and 0.08 ms, respectively, while the inference stage requires just 0.10 ms per image.
In the inference stage, the system with CR = $2/24$ is capable of processing 10,000 frames per second, showcasing its strong real-time communication capabilities.
Additionally, when the compression ratio increases to $4/24$, the processing time per image in both the training and inference stages increases by only about 0.02 ms. This demonstrates that increasing the compression ratio has a negligible effect on inference speed, further highlighting the system's impressive efficiency. 
The low inference time of our proposed system highlights its scalability, allowing it to be easily extended to more complex tasks, such as real-time video transmission.

\subsection{The Performance Comparison on Security of Different Approaches}
\begin{figure}[htbp]
\begin{center}
\centerline{\includegraphics[width=1\linewidth]{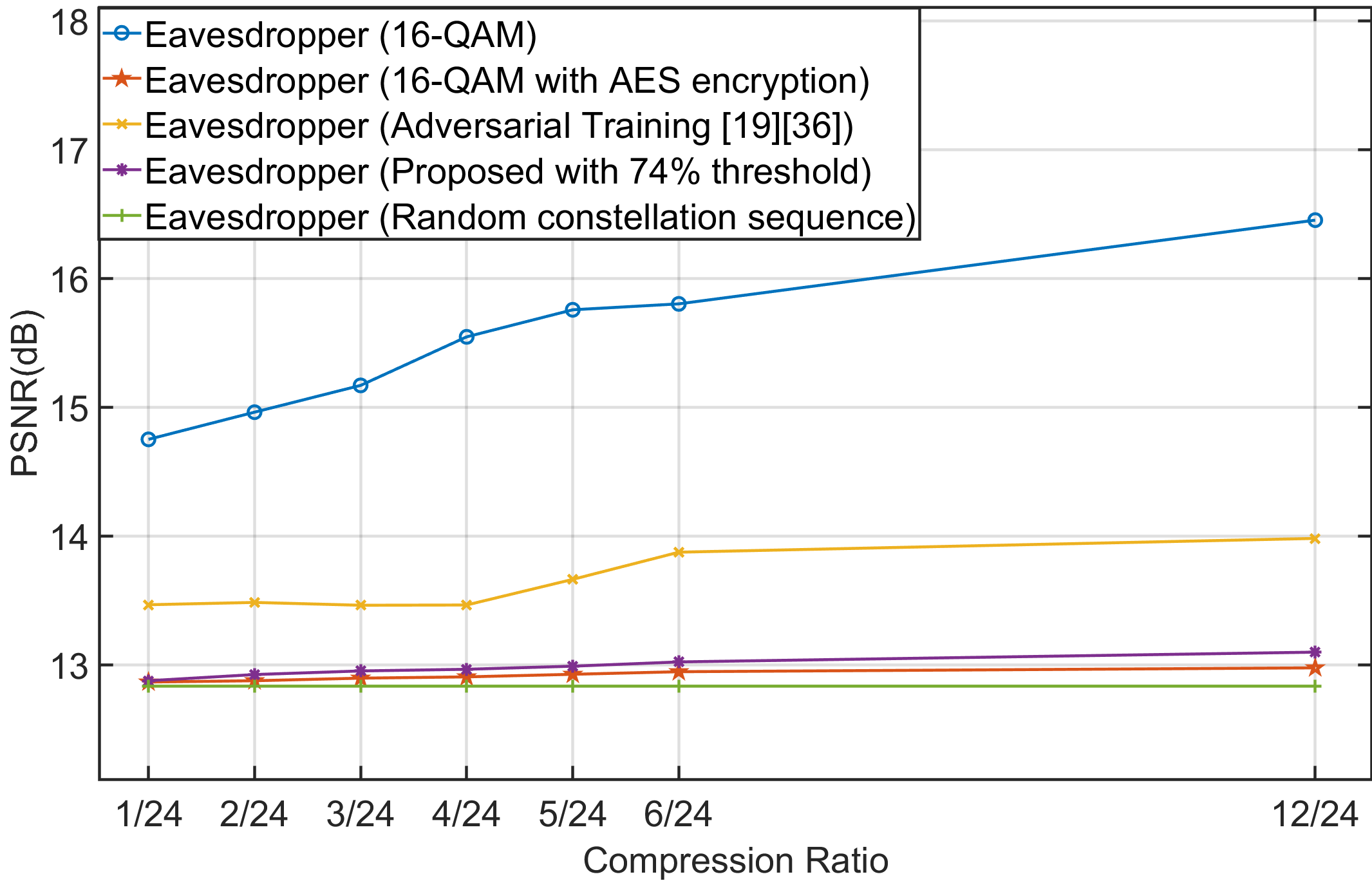}}
\caption{The PSNR performance of the eavesdropper by different approaches. Here, $\mathrm{SNR}_{\mathrm{leg}}=20\mathrm{dB}$ and $\mathrm{SNR}_{\mathrm{eve}}=-10\mathrm{dB}$.}
\label{fig.plot3}
\end{center}
\vskip -0.2in
\end{figure}

In this subsection, we assess the security of our proposed secure digital SemCom system in comparison to benchmarks. Specifically, we first examine the PSNR performance of the eavesdropper, where a lower PSNR indicates a more secure system. Fig.~\ref{fig.plot3} illustrates the PSNR performance of the eavesdropper across different methods and compression ratios. 
The channel conditions are configured with $\mathrm{SNR}_{\mathrm{leg}}=20\mathrm{dB}$ and $\mathrm{SNR}_{\mathrm{eve}}=-10\mathrm{dB}$. In our proposed system, we set the eavesdropper's SEP to be 74\%, corresponding to (MSEP) $b=74\%$ and, consequently, the power allocation coefficient (PAC) $a=0.014$.

\begin{figure}[htbp]
\begin{center}
\centerline{\includegraphics[width=1\linewidth]{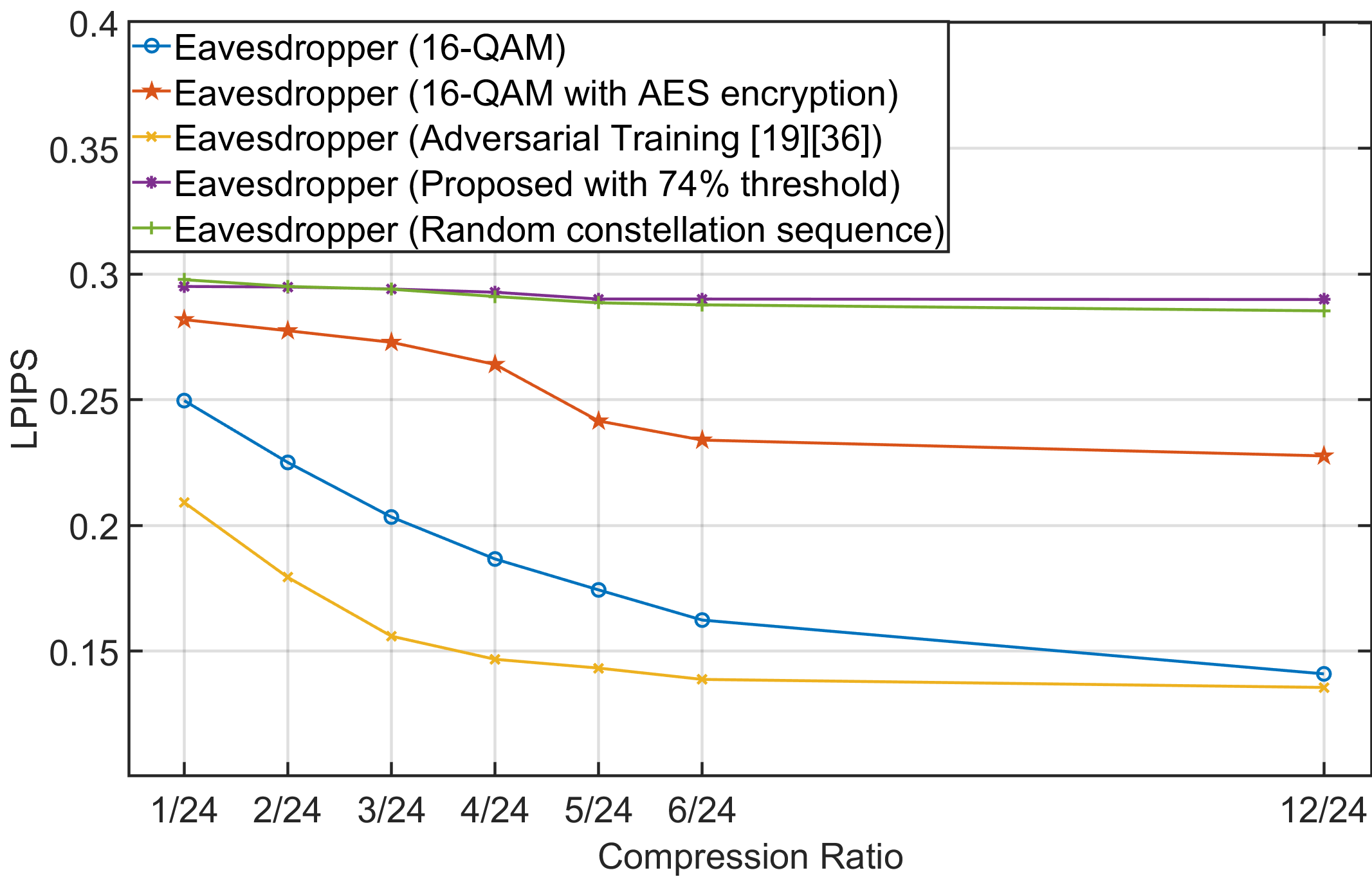}}
\caption{The LPIPS performance of the eavesdropper by different approaches. Here, $\mathrm{SNR}_{\mathrm{leg}}=20\mathrm{dB}$ and $\mathrm{SNR}_{\mathrm{eve}}=-10\mathrm{dB}$.}
\label{fig.plot3-lpips}
\end{center}
\vskip -0.3in
\end{figure}

\begin{figure}[htbp]
\begin{center}
\centerline{\includegraphics[width=1\linewidth]{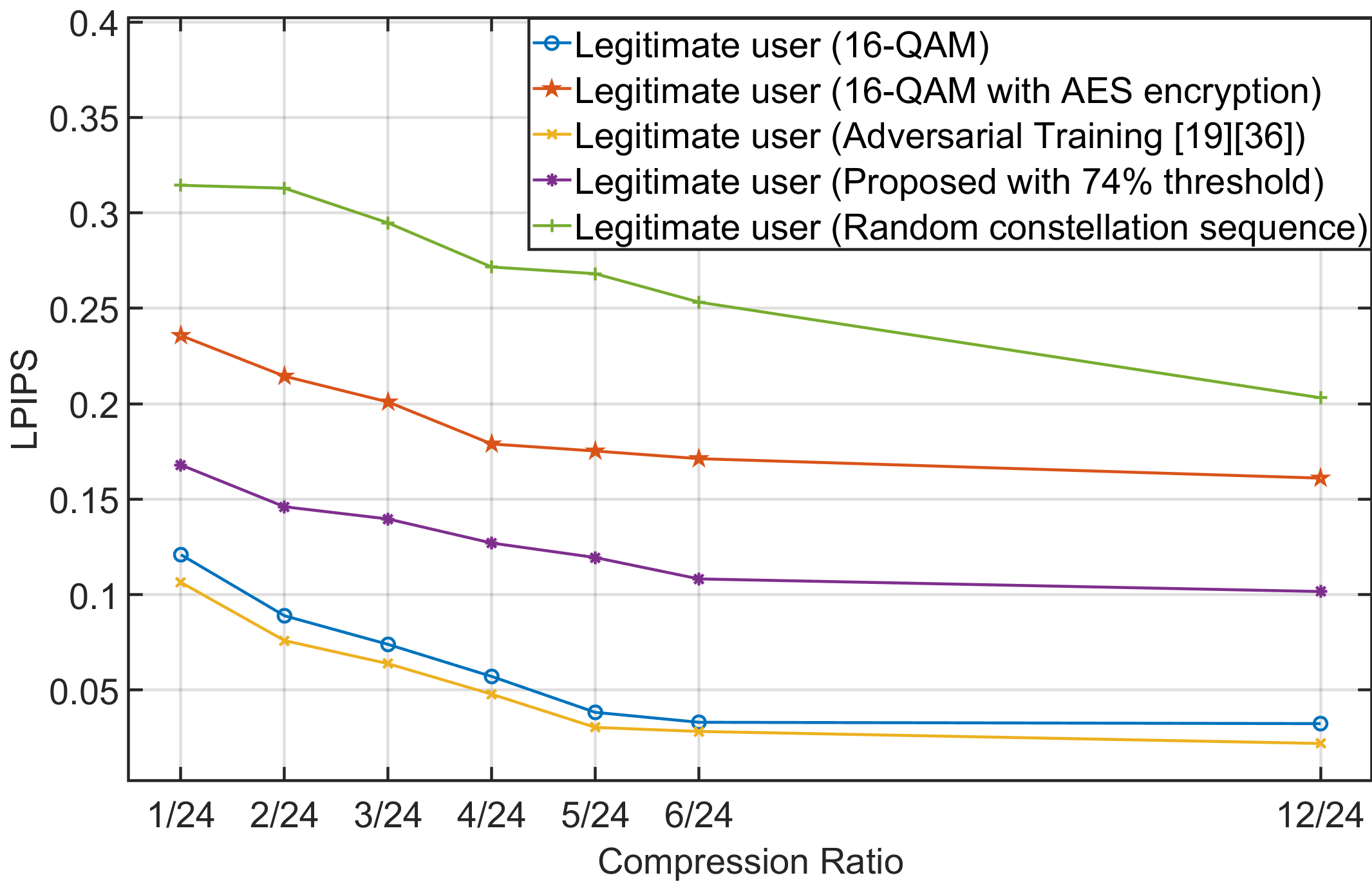}}
\caption{The LPIPS performance of the legitimate user by different approaches. Here, $\mathrm{SNR}_{\mathrm{leg}}=20\mathrm{dB}$ and $\mathrm{SNR}_{\mathrm{eve}}=-10\mathrm{dB}$.}
\label{fig.plot3-lpips-legitimate}
\end{center}
\vskip -0.3in
\end{figure}
 
From Fig.~\ref{fig.plot3}, it is evident that the security of our proposed system surpasses both \textit{16-QAM without Superposition} and \textit{Adversarial Training}. \textit{16-QAM without Superposition}, lacking specific security guarantee, serves as a baseline to gauge the minimum security, highlighting the improvement achieved by other methods.
The numerical results reveal two main security disadvantages in \textit{16-QAM without Superposition}. Firstly, the only obstacle for the eavesdropper to obtain semantic information is the high-power AWGN and the low compression ratio. This is not sufficient to prevent the eavesdropper from reconstructing the original image data with a relatively high PSNR. For instance, at a compression ratio of $12/24$, the PSNR of the recovered image can reach 16.45dB, surpassing \textit{Adversarial Training} by 2.47dB and our proposed system by 3.35dB.
Secondly, information leakage increases rapidly with the growth of the compression ratio, rendering the security of the SemCom system extremely vulnerable in the high compression rate regime. For instance, at a compression ratio of $12/24$, the PSNR of the reconstructed image rises to 16.45dB, which is 1.7dB higher than when the compression ratio is $1/24$. These findings underscore the importance of designing secure SemCom systems.

The security enhancement of \textit{Adversarial Training} is notable compared to \textit{16-QAM without Superposition}. This improvement is attributed to the strategic engagement between the legitimate user and the eavesdropper, allowing the former to minimize information leakage actively. Consequently, there is a significant reduction in the PSNR performance of the eavesdropper compared to \textit{16-QAM without Superposition}. For instance, at a compression ratio of $1/24$, \textit{Adversarial Training} lowers the PSNR performance of the eavesdropper by 1.28dB, and at a compression ratio of $12/24$, it reduces the PSNR performance by 2.47dB. This demonstrates the impact of actively defending against the eavesdropper in improving SemCom system security. However, it's worth noting that the security of \textit{Adversarial Training} diminishes notably at high compression ratios. For instance, there is a 0.4-0.5dB improvement in the PSNR performance of the eavesdropper as the compression rate surpasses $4/24$. This implies that, while \textit{Adversarial Training} proves effective at low compression rates, its security diminishes at relatively high compression rates due to increased information leakage.

In contrast, our proposed system exhibits superior security. For instance, in our system, the PSNR performance of the eavesdropper is reduced by 0.6-0.9dB compared to \textit{Adversarial Training}. Furthermore, with an increase in compression rate from $1/24$ to $12/24$, the PSNR performance of the eavesdropper improves by only 0.2dB. It's essential to highlight that the PSNR performance of the eavesdropper in our proposed system is approximately 12.9-13.1dB, which is nearly equivalent to the scenario where the eavesdropper receives a random constellation sequence. 
Additionally, the PSNR performance of the eavesdropper in our proposed system closely matches that of \textit{16-QAM with AES Encryption}, indicating that the security of our approach is almost equivalent to that of classical encryption methods.
Moreover, our proposed system outperforms \textit{16-QAM with AES Encryption} in terms of inference speed. 
For instance, at a CR of $2/24$, the average inference time per image in \textit{16-QAM with AES Encryption} is 0.02 ms longer than in our proposed system, and at a CR of $4/24$, it is 0.04 ms longer.
This indicates that the eavesdropper is unable to recover much valid information from the received signals, demonstrating that the security of the proposed approach is close to the experimental upper bound.

\begin{figure*}[htbp]
\centering
\subfigure[Source image]{
\begin{minipage}[t]{0.24\linewidth}
\centering
\includegraphics[width=0.9in]{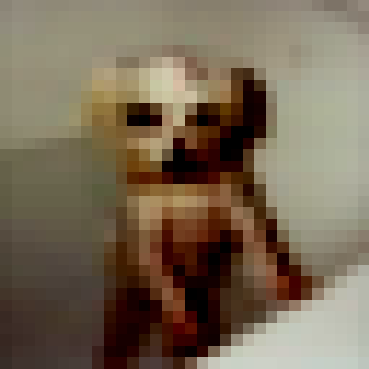}
\end{minipage}%
}%
\subfigure[Our proposed system]{
\begin{minipage}[t]{0.24\linewidth}
\centering
\includegraphics[width=0.9in]{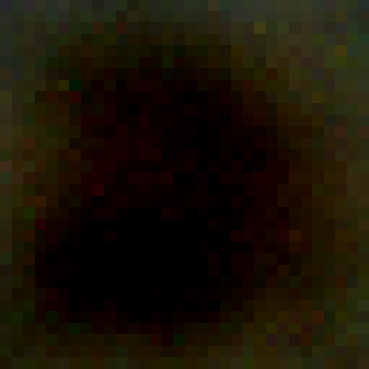}
\end{minipage}%
}%
\subfigure[\textit{16-QAM}]{
\begin{minipage}[t]{0.24\linewidth}
\centering
\includegraphics[width=0.9in]{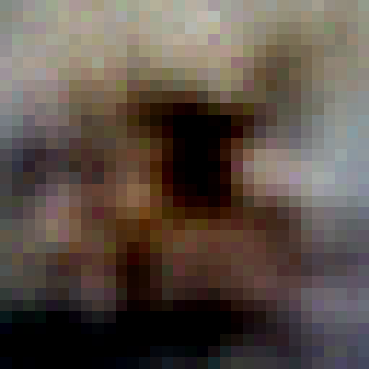}
\end{minipage}
}%
\subfigure[\textit{Adversarial Training}]{
\begin{minipage}[t]{0.24\linewidth}
\centering
\includegraphics[width=0.9in]{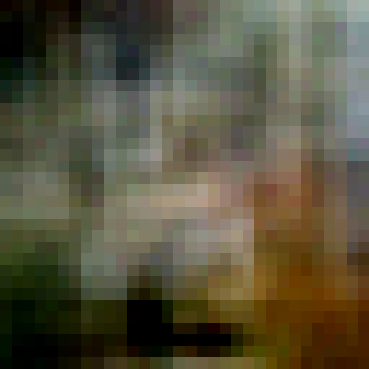}
\end{minipage}
}%
\centering
\caption{Visual analysis of the reconstructed images at the eavesdropper. The channel conditions are set at $\mathrm{SNR}_{\mathrm{leg}}=20\mathrm{dB}$ and $\mathrm{SNR}_{\mathrm{eve}}=-10\mathrm{dB}$. The compression ratio is $6/24$.}
\label{fig.visual}
\vskip -0.2in
\end{figure*}

We then evaluate the LPIPS performance of the eavesdropper, where a higher LPIPS score indicates better semantic security for the system.
The experimental setup follows that shown in Fig.~\ref{fig.plot3}.
Fig.~\ref{fig.plot3-lpips} illustrates the LPIPS performance of the eavesdropper across different methods and compression ratios.
It is clear that our proposed method achieves nearly the same level of semantic security as the \textit{Random Constellation Sequence} benchmark, 
further supporting that our approach is very close to achieving the experimental upper bound of security.
This implies that the eavesdropper's recovered image has very poor perceptual fidelity, indicating that almost no meaningful information has been obtained. 
In contrast, the LPIPS values of the eavesdropper in the other methods are relatively lower, indicating that these methods provide weaker security.
%
We also evaluate the reliability of our proposed system in terms of semantic task performance. 
Specifically, we assess the perceptual quality of the reconstructed images at the legitimate user using the LPIPS metric, as shown in Fig.~\ref{fig.plot3-lpips-legitimate}. 
Fig.~\ref{fig.plot3-lpips-legitimate} presents the LPIPS performance of the legitimate user across different methods and compression ratios. 
We observe that our proposed method substantially outperforms \textit{16-QAM with AES Encryption} in terms of semantic task performance, although it falls short compared to \textit{Adversarial Training} and the unprotected \textit{16-QAM without Superposition}.
We note that AES encryption increases LPIPS compared to plain 16-QAM, due to error propagation across multiple blocks caused by CBC mode under noisy channels.
In conjunction with Fig.~\ref{fig.plot3-lpips}, which shows that both \textit{Adversarial Training} and \textit{16-QAM without Superposition} perform very poorly in terms of semantic security, our proposed method achieves the best semantic security among the evaluated approaches.
In summary, our proposed method achieves high semantic security with a slight reduction in semantic task performance. It clearly outperforms \textit{16-QAM with AES Encryption} in both reliability and semantic security, and demonstrates the strongest semantic security among all benchmarks.

In Fig.~\ref{fig.visual}, we visually compare the reconstructed images at the eavesdropper by the proposed system and the benchmarks.
Fig.~\ref{fig.visual}(a) is the source image $\textbf{X}$, Fig.~\ref{fig.visual}(b), Fig.~\ref{fig.visual}(c) and Fig.~\ref{fig.visual}(d) show the recovered images by the eavesdropper in our proposed system, \textit{16-QAM without Superposition}, and \textit{Adversarial Training}, respectively.
Comparing the recovered images of the eavesdropper, we demonstrate that the security of our proposed system is better than the two benchmarks. The source image, in this case, is a picture of a dog. In the \textit{16-QAM without Superposition} benchmark, the eavesdropper can recover a clear outline of the source image. In \textit{Adversarial Training}, the eavesdropper can only recover some blurred background. However, in our proposed system, the recovered image of the eavesdropper is nearly completely black, making it extremely challenging for the eavesdropper to obtain any meaningful information. Therefore, our proposed system excels over the benchmarks in preventing information leakage to the eavesdropper.


\subsection{Performance of Our Proposed System under Different SEPs of the Eavesdropper}

\begin{figure}[htbp]
\begin{center}
\centerline{\includegraphics[width=1\linewidth]{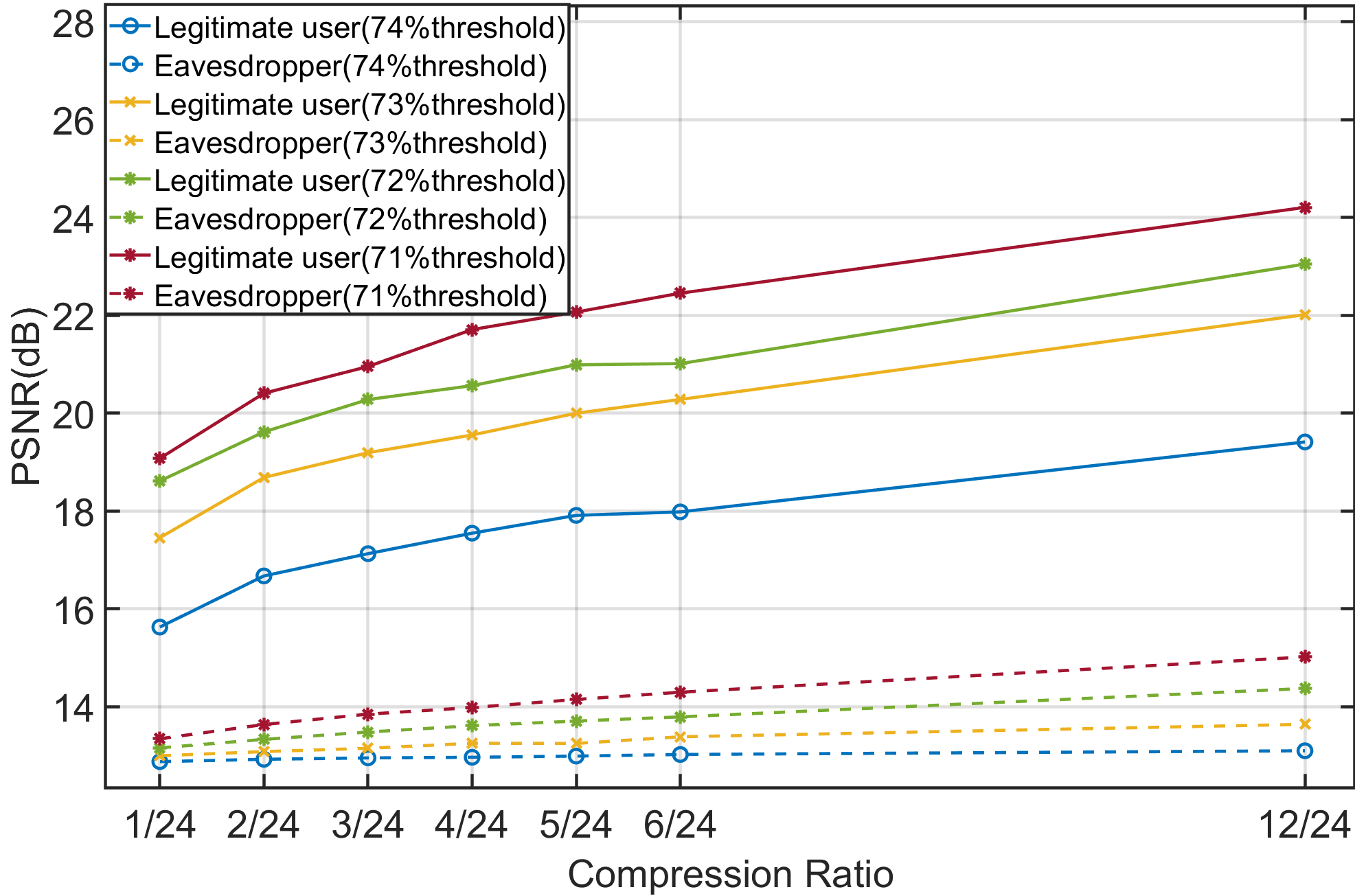}}
\caption{PSNR performance under different SEPs of the eavesdropper, i.e., different MSEPs. Here, $\mathrm{SNR}_{\mathrm{leg}}=20\mathrm{dB}$ and $\mathrm{SNR}_{\mathrm{eve}}=-10\mathrm{dB}$.}
\label{fig.plot1}
\end{center}
\vskip -0.2in
\end{figure}

In this subsection, we evaluate the performance and security of the proposed system under different MSEPs, aiming to analyze the impact of MSEPs on both aspects. 
Specifically, we evaluate the PSNR performance of both the legitimate user and the eavesdropper, as illustrated in Fig.~\ref{fig.plot1}. The channel conditions are configured with $\mathrm{SNR}_{\mathrm{leg}}=20\mathrm{dB}$ and $\mathrm{SNR}_{\mathrm{eve}}=-10\mathrm{dB}$. 
The MSEPs vary from 71\% to 74\% in 1\% increments, where a lower MSEP corresponds to a larger PAC. 
Note that each model is trained from scratch based on the specified MSEP.
To represent the PSNR performance at the same MSEP, consistent color curves are used. The solid line depicts the PSNR performance of the legitimate user, while the dashed line represents the PSNR performance of the eavesdropper.

From Fig.~\ref{fig.plot1}, it is evident that a lower MSEP leads to better PSNR performance for both the legitimate user and the eavesdropper. This is because, as the SEP of the eavesdropper decreases, the SEP of the legitimate user also decreases. Specifically, the system with a 71\% MSEP exhibits the best PSNR performance for the legitimate user but has the weakest security, as the PSNR performance of the eavesdropper is relatively high at this point. As MSEP increases, the PSNR performance of the legitimate user decreases while its security improves. The system with a 74\% MSEP demonstrates the highest security, making it nearly impossible for the eavesdropper to recover any useful information. In our proposed system, the PSNR performance of the eavesdropper improves gradually as the compression ratio increases. For the system with a 71\% MSEP, as the compression ratio increases from $1/24$ to $12/24$, the PSNR performance of the eavesdropper improves by 1.7dB. In contrast, for the system with a 74\% MSEP, the PSNR performance of the eavesdropper improves by only 0.2dB. Therefore, to achieve high security, a MSEP of 74\% should be employed to minimize information leakage. However, if high security is not a strict requirement, MSEP can be appropriately lowered to enhance the PSNR performance of the legitimate user.

Following the experimental setting in Fig.~\ref{fig.plot1}, we calculate the mutual information between the outer constellation sequence $\textbf{Y}_1$ and the constellation sequence decoded by the eavesdropper $\bar{\textbf{Z}}_2$ in our proposed system, as illustrated in Table~\ref{tab2}. 
To calculate the mutual information between $\textbf{Y}_1$ and $\bar{\textbf{Z}}_2$, we employ the mutual information neural estimation (MINE) method proposed by Belghazi \textit{et al.} \cite{belghazi2018mutual}.
This method estimates mutual information by maximizing a dual representation of the Kullback-Leibler (KL) divergence between the joint distribution and the product of the marginal distributions.
Specifically, MINE frames this estimation as a maximization problem, where an NN is trained to approximate the KL divergence. The network learns to distinguish between the joint distribution of the variables and the product of their marginals, capturing the degree of dependency between them. This approach allows for accurate and efficient estimation of mutual information, which is crucial for analyzing the security of our proposed system.
%
We observe that the mutual information values are consistently close to 0 across all MSEPs. This suggests that the eavesdropper can hardly obtain any information about the source image, thereby confirming that the security of our proposed system has nearly reached the experimental upper bound.
As MSEP increases, the mutual information value decreases, indicating that adjusting MSEP can control the security of the system, and higher MSEPs imply higher security.

\begin{table}[htbp]
\caption{The mutual information between $\textbf{Y}_1$ and $\bar{\textbf{Z}}_2$ under different MSEPs. The compression ratio is set to $6/24$.}
\centering
\begin{tabular}{c|c|c|c|c}
\hline
MSEP (\%) & 71   & 72   & 73   & 74  \\ \hline
Mutual Information & 0.017 & 0.012 & 0.009 & 0.006 \\ \hline
\end{tabular}
\label{tab2}
\end{table}

\begin{figure*}[htbp]
\centering
\subfigure[MSEP=74\%]{
\begin{minipage}[t]{0.24\linewidth}
\centering
\includegraphics[width=0.9in]{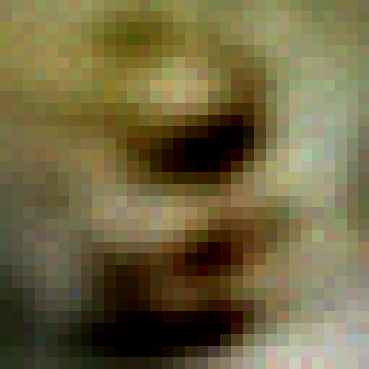}
\end{minipage}%
}%
\subfigure[MSEP=73\%]{
\begin{minipage}[t]{0.24\linewidth}
\centering
\includegraphics[width=0.9in]{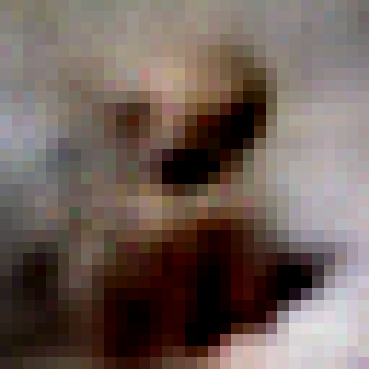}
\end{minipage}%
}%
\subfigure[MSEP=72\%]{
\begin{minipage}[t]{0.24\linewidth}
\centering
\includegraphics[width=0.9in]{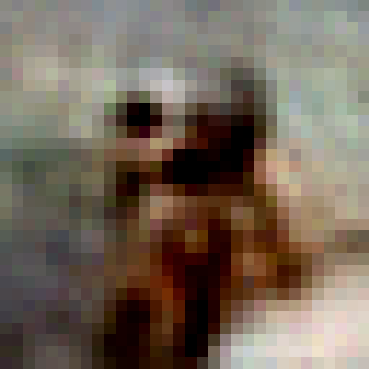}
\end{minipage}
}%
\subfigure[MSEP=71\%]{
\begin{minipage}[t]{0.24\linewidth}
\centering
\includegraphics[width=0.9in]{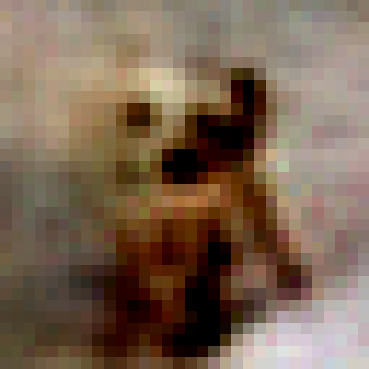}
\end{minipage}
}%
\centering
\caption{Visual analysis of the reconstructed images by the legitimate user under different MSEPs. The channel conditions are set at $\mathrm{SNR}_{\mathrm{leg}}=20\mathrm{dB}$ and $\mathrm{SNR}_{\mathrm{eve}}=-10\mathrm{dB}$. The compression ratio is $12/24$.}
\label{fig.visual2}
\vskip -0.2in
\end{figure*}

\begin{figure*}[htbp]
\centering
\subfigure[MSEP=74\%]{
\begin{minipage}[t]{0.24\linewidth}
\centering
\includegraphics[width=0.9in]{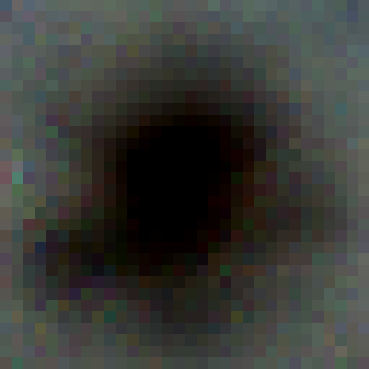}
\end{minipage}%
}%
\subfigure[MSEP=73\%]{
\begin{minipage}[t]{0.24\linewidth}
\centering
\includegraphics[width=0.9in]{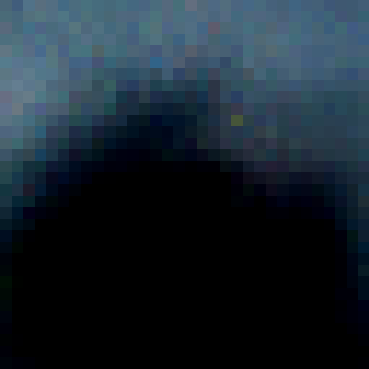}
\end{minipage}%
}%
\subfigure[MSEP=72\%]{
\begin{minipage}[t]{0.24\linewidth}
\centering
\includegraphics[width=0.9in]{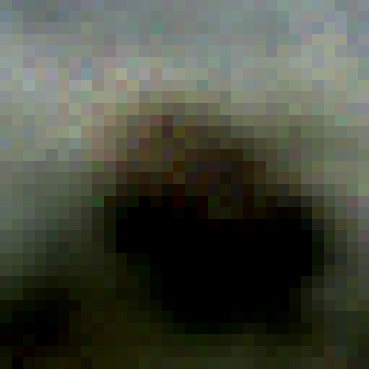}
\end{minipage}
}%
\subfigure[MSEP=71\%]{
\begin{minipage}[t]{0.24\linewidth}
\centering
\includegraphics[width=0.9in]{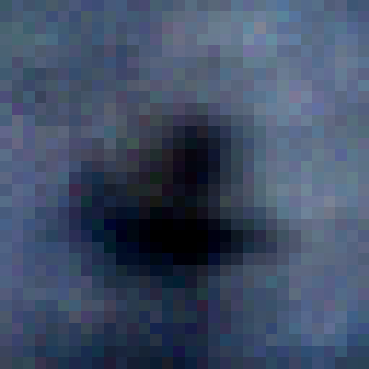}
\end{minipage}
}%
\centering
\caption{Visual analysis of the reconstructed images by the eavesdropper under different MSEPs. The channel conditions are set at $\mathrm{SNR}_{\mathrm{leg}}=20\mathrm{dB}$ and $\mathrm{SNR}_{\mathrm{eve}}=-10\mathrm{dB}$. The compression ratio is $12/24$.}
\label{fig.visual3}
\vskip -0.1in
\end{figure*}

\begin{figure}[htbp]
\begin{center}
\centerline{\includegraphics[width=1\linewidth]{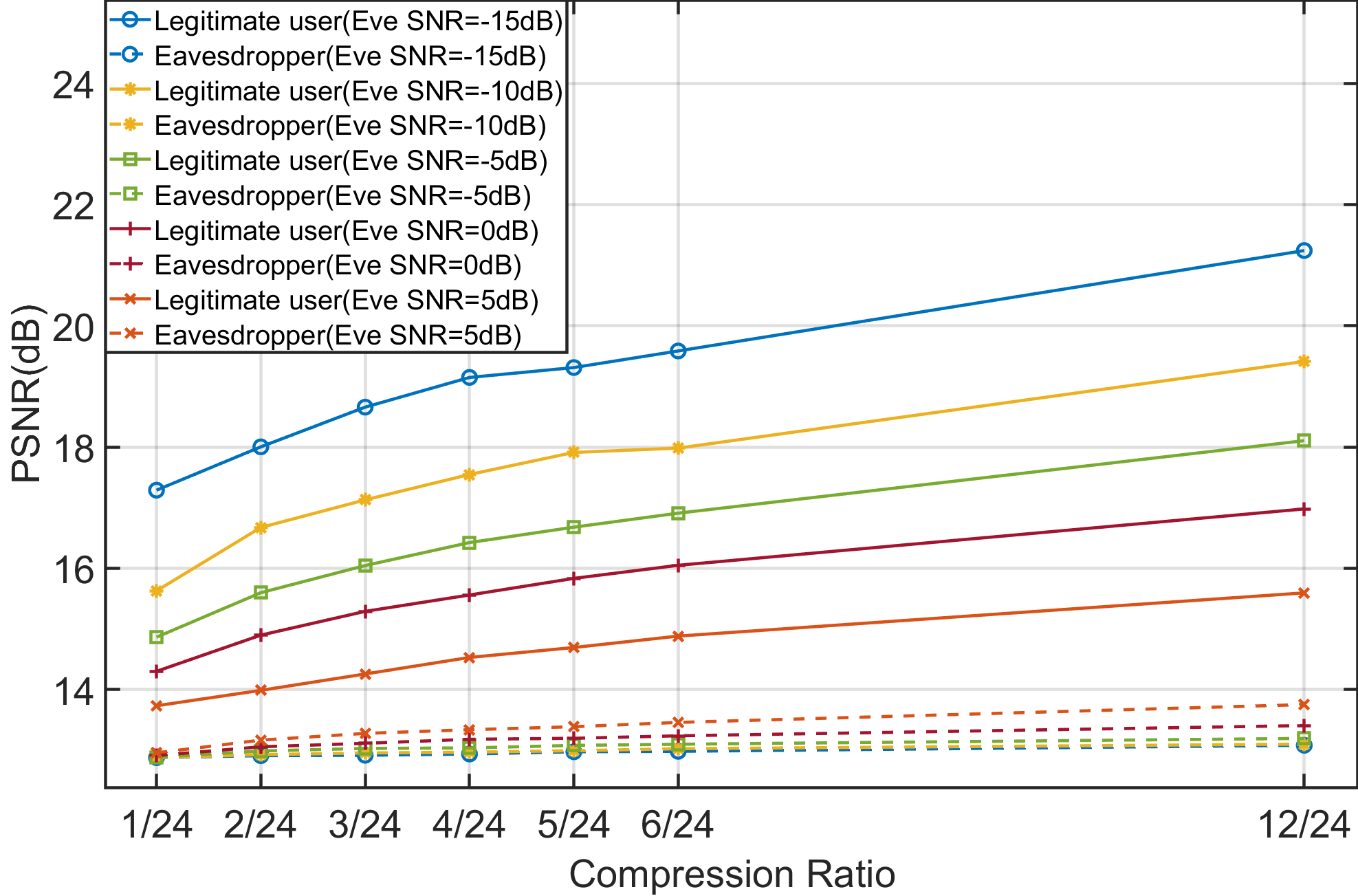}}
\caption{PSNR performance under different legitimate user-eavesdropper SNR pairs. Here MSEP is set to 74\% and $\mathrm{SNR}_{\mathrm{leg}}=20\mathrm{dB}$.}
\label{fig.plot17}
\end{center}
\vskip -0.3in
\end{figure}

We also present the images reconstructed by the legitimate user and the eavesdropper at various MSEPs, as illustrated in Fig.~\ref{fig.visual2} and Fig.~\ref{fig.visual3}, respectively. In Fig.~\ref{fig.visual2}, it is observed that the reconstructed image by the legitimate user becomes more distinct as MSEP decreases. Fig.~\ref{fig.visual2}(a) only provides a basic outline of the source image, while Fig.~\ref{fig.visual2}(d) manages to capture more details of the source image. On the other hand, Fig.~\ref{fig.visual3} reveals that even with a lowered MSEP, the reconstructed image by the eavesdropper contains minimal useful information and remains largely blurred.

\subsection{Performance of Our Proposed System under Different Legitimate user-Eavesdropper SNR Pairs}

In this subsection, we assess the performance of the proposed system across diverse channel conditions, as illustrated in Fig.~\ref{fig.plot17}, where the channel conditions are set at $\mathrm{SNR}_{\mathrm{leg}}=20\mathrm{dB}$ and $\mathrm{SNR}_{\mathrm{eve}}\in \{-15,-10,-5,0,5\}\mathrm{dB}$. 
Note that each model is trained from scratch based on the specified legitimate user-eavesdropper SNR pair.
With a fixed MSEP, we observe that the SEP of the legitimate user decreases as the difference between the legitimate channel SNR $\mathrm{SNR}_{\mathrm{leg}}$ and the eavesdropper channel SNR $\mathrm{SNR}_{\mathrm{eve}}$ increases. We can observe that with $\mathrm{SNR}_{\mathrm{leg}}$ at 20dB and MSEP at 74\%, the PSNR performance of the legitimate user decreases with the eavesdropper channel SNR $\mathrm{SNR}_{\mathrm{eve}}$. Conversely, the PSNR performance of the eavesdropper shows only marginal improvement. Even with an elevation in the eavesdropper channel SNR $\mathrm{SNR}_{\mathrm{eve}}$ from -15dB to 5dB, the improvement in the PSNR performance of the eavesdropper ranges from 0.1 to 0.7dB. Moreover, as the compression rate increases, there is also slight enhancement in the PSNR performance of the eavesdropper. These findings indicate that even with improvements in the eavesdropper's channel conditions or an increase in the number of acquired symbols, our proposed system ensures that the eavesdropper is nearly incapable of recovering the source image, thereby approaching the experimental upper bound of security.
%
We infer that further increases in $\mathrm{SNR}_{\mathrm{eve}}$ (e.g., to 10dB or higher) may still compromise system security, even with a 74\% MSEP, suggesting the need for a higher MSEP. Conversely, when $\mathrm{SNR}_{\mathrm{eve}}$ is low (e.g., -15dB) and the compression rate is low (e.g., $1/24$), good security can be achieved even with a MSEP value of 71\%. This validates the system's adaptability to different application scenarios, allowing us to tailor MSEPs for optimal performance and high security under various channel conditions.

\begin{figure}[htbp]
\begin{center}
\centerline{\includegraphics[width=1\linewidth]{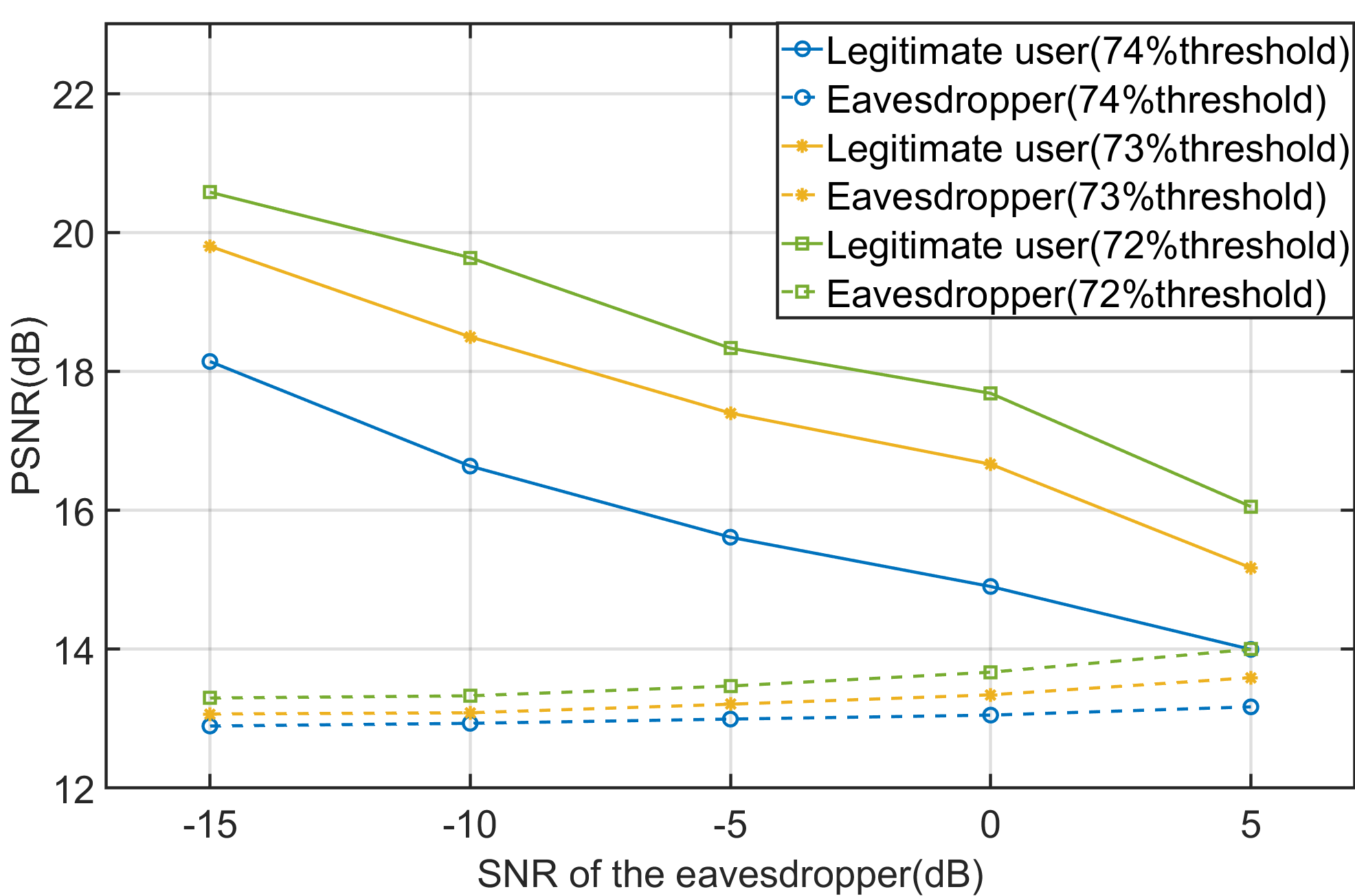}}
\caption{PSNR performance under different MSEPs and channel conditions. Here the compression ratio is fixed at $2/24$ and $\mathrm{SNR}_{\mathrm{leg}}=20\mathrm{dB}$.}
\label{fig.plotmsep}
\end{center}
\vskip -0.3in
\end{figure}

\subsection{Performance of Our Proposed System under Different SEPs and SNRs of the Eavesdropper}


In this subsection, we evaluate the performance of the proposed system across various MSEPs and channel conditions, as shown in Fig.~\ref{fig.plotmsep}. 
The channel conditions are set with $\mathrm{SNR}_{\mathrm{leg}}=20\mathrm{dB}$ and $\mathrm{SNR}_{\mathrm{eve}}\in \{-15,-10,-5,0,5\}\mathrm{dB}$, 
with the compression ratio fixed at $2/24$. 
The MSEPs range from 72\% to 74\% in 1\% increments.
As shown in Fig.~\ref{fig.plotmsep}, when the MSEP is fixed, the PSNR of the legitimate user decreases as the eavesdropper's channel quality improves, while the eavesdropper's PSNR increases very slowly with the improvement of its channel condition. 
This trend is consistent with our SEP derivation for both the legitimate user and the eavesdropper.
As the SNR gap between the legitimate user and the eavesdropper narrows, the legitimate user may need to sacrifice some performance to maintain security. 
When the MSEP is set to 74\%, the eavesdropper's PSNR shows little improvement, but in this maximum security setting, the legitimate user experiences a significant drop in performance.
Furthermore, as the MSEP decreases, the PSNR of the legitimate user increases, effectively trading some security for better performance.
Specifically, when the MSEP is reduced to 72\%, the eavesdropper's PSNR increases by only about 0.4-0.8dB compared to when the MSEP is 74\%, while the legitimate user's PSNR improves by 0.8-2.4dB. 
In this case, the increase in the eavesdropper's PSNR is marginal. Therefore, if strict security is not a requirement, a slight reduction in MSEP can lead to a significant improvement in the performance of the legitimate user. This trade-off demonstrates that a small sacrifice in security can yield considerable benefits in terms of performance for the legitimate user. Additionally, increasing the legitimate user's SNR advantage over the eavesdropper can further enhance security.
Overall, when balancing security and legitimate user performance requirements, the proposed power allocation scheme provides a flexible approach, allowing for the selection of an optimal solution based on specific needs. This flexibility highlights the contribution of the proposed approach in this paper.
%

\subsection{Performance of Our Proposed System under Underestimated Eavesdropper SNR}

\begin{figure}[htbp]
\begin{center}
\centerline{\includegraphics[width=1\linewidth]{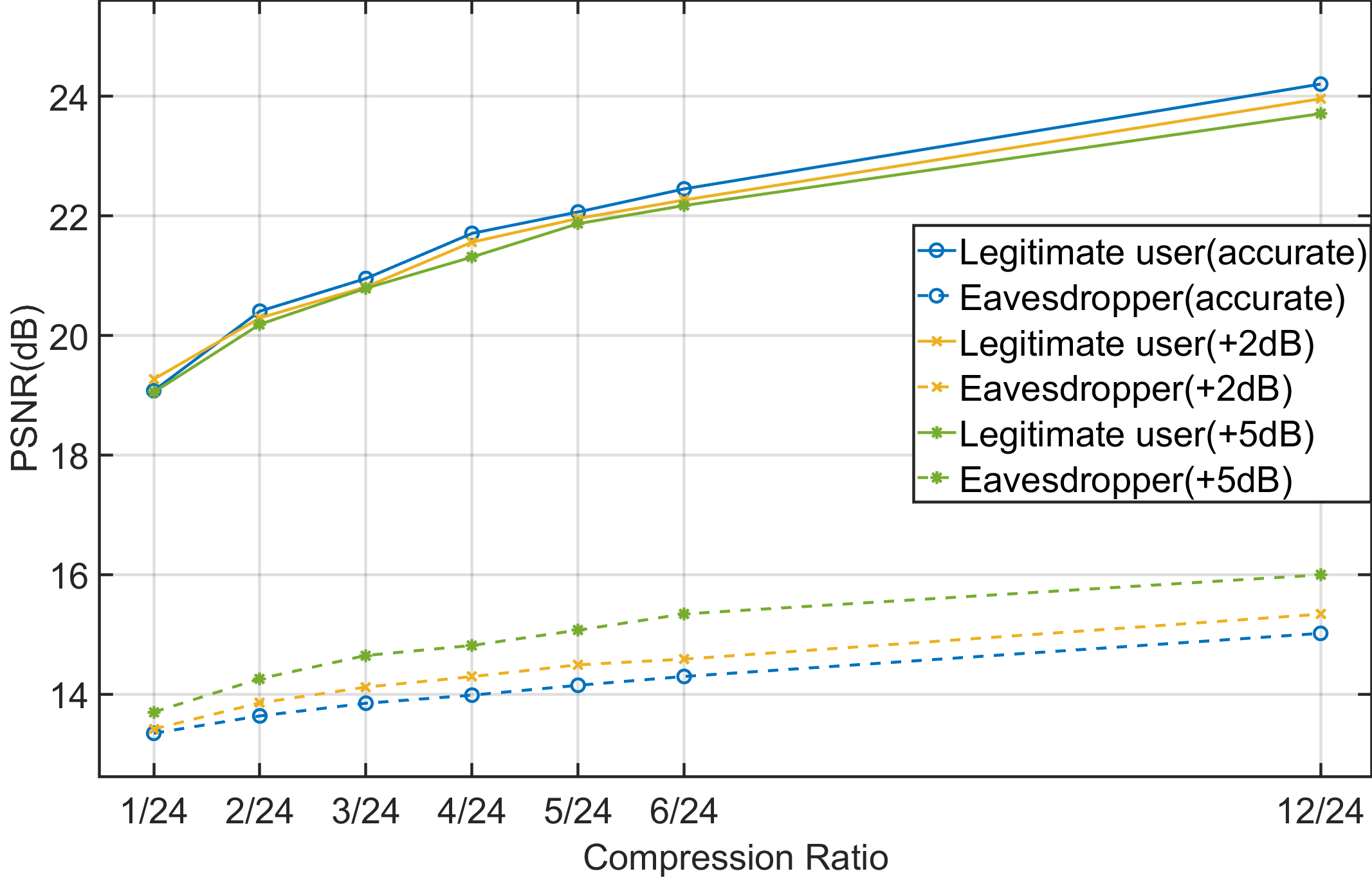}}
\caption{PSNR performance of the proposed system under underestimated eavesdropper SNR. 
The expected SNR values are $\mathrm{SNR}_{\mathrm{leg}}=20\mathrm{dB}$ and $\mathrm{SNR}_{\mathrm{eve}}=-10\mathrm{dB}$, with the MSEP set to 71\%. 
``accurate'' refers to the case where $\mathrm{SNR}_{\mathrm{eve}}$ matches the expected value, while ``+2dB'' and ``+5dB'' indicate that the actual $\mathrm{SNR}_{\mathrm{eve}}$ is 2dB and 5dB higher than expected, respectively.
}
\label{fig.plot_higher_eve_snr}
\end{center}
\vskip -0.3in
\end{figure}

\begin{figure}[htbp]
\begin{center}
\centerline{\includegraphics[width=1\linewidth]{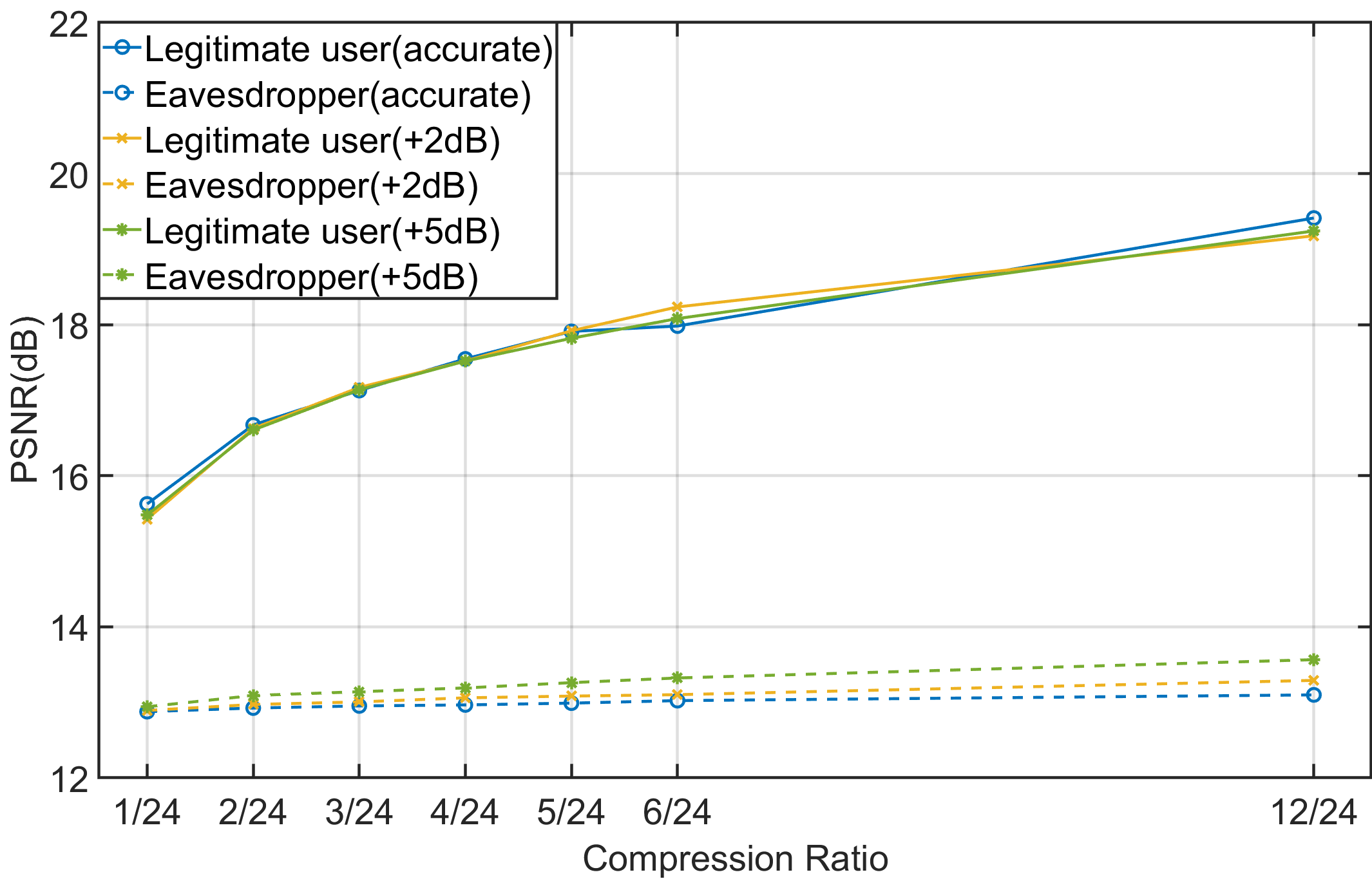}}
\caption{PSNR performance of the proposed system under underestimated eavesdropper SNR. 
The expected SNR values are $\mathrm{SNR}_{\mathrm{leg}}=20\mathrm{dB}$ and $\mathrm{SNR}_{\mathrm{eve}}=-10\mathrm{dB}$, with the MSEP set to 74\%. 
%
%
}
\label{fig.plot_higher_eve_snr_MSEP74}
\end{center}
\vskip -0.3in
\end{figure}

In this subsection, we analyze the performance of our proposed system when Alice and Bob have inaccurate knowledge of Eve's SNR, with the MSEP set to 71\% and 74\%, respectively, as shown in Fig.~\ref{fig.plot_higher_eve_snr} and Fig.~\ref{fig.plot_higher_eve_snr_MSEP74}. 
%
Specifically, we assume that Alice and Bob expect $\mathrm{SNR}_{\mathrm{eve}}=-10\mathrm{dB}$, while in practice the actual $\mathrm{SNR}_{\mathrm{eve}}$ may be 2dB or 5dB higher.
$\mathrm{SNR}_{\mathrm{leg}}$ is fixed at 20dB, and the MSEP is set to 71\% or 74\%.
From Fig.~\ref{fig.plot_higher_eve_snr}, we observe that as the actual $\mathrm{SNR}_{\mathrm{eve}}$ increases beyond the expected value, the PSNR performance of the eavesdropper slightly improves. In contrast, the PSNR performance of the legitimate user remains largely unaffected. 
From Fig.~\ref{fig.plot_higher_eve_snr_MSEP74}, similar observations can be made.
%
These results demonstrate that the proposed system maintains strong robustness against inaccuracies in the knowledge of Eve's SNR. 
While the security level decreases slightly with increasing $\mathrm{SNR}_{\mathrm{eve}}$, the task performance remains stable.
Moreover, in scenarios where Eve's SNR is uncertain, our proposed system can be designed based on the maximum expected $\mathrm{SNR}_{\mathrm{eve}}$, thereby ensuring a reliable security guarantee.


\subsection{Performance of Our Proposed System with and without Superposition Coding}

\begin{figure}[htbp]
\begin{center}
\centerline{\includegraphics[width=1\linewidth]{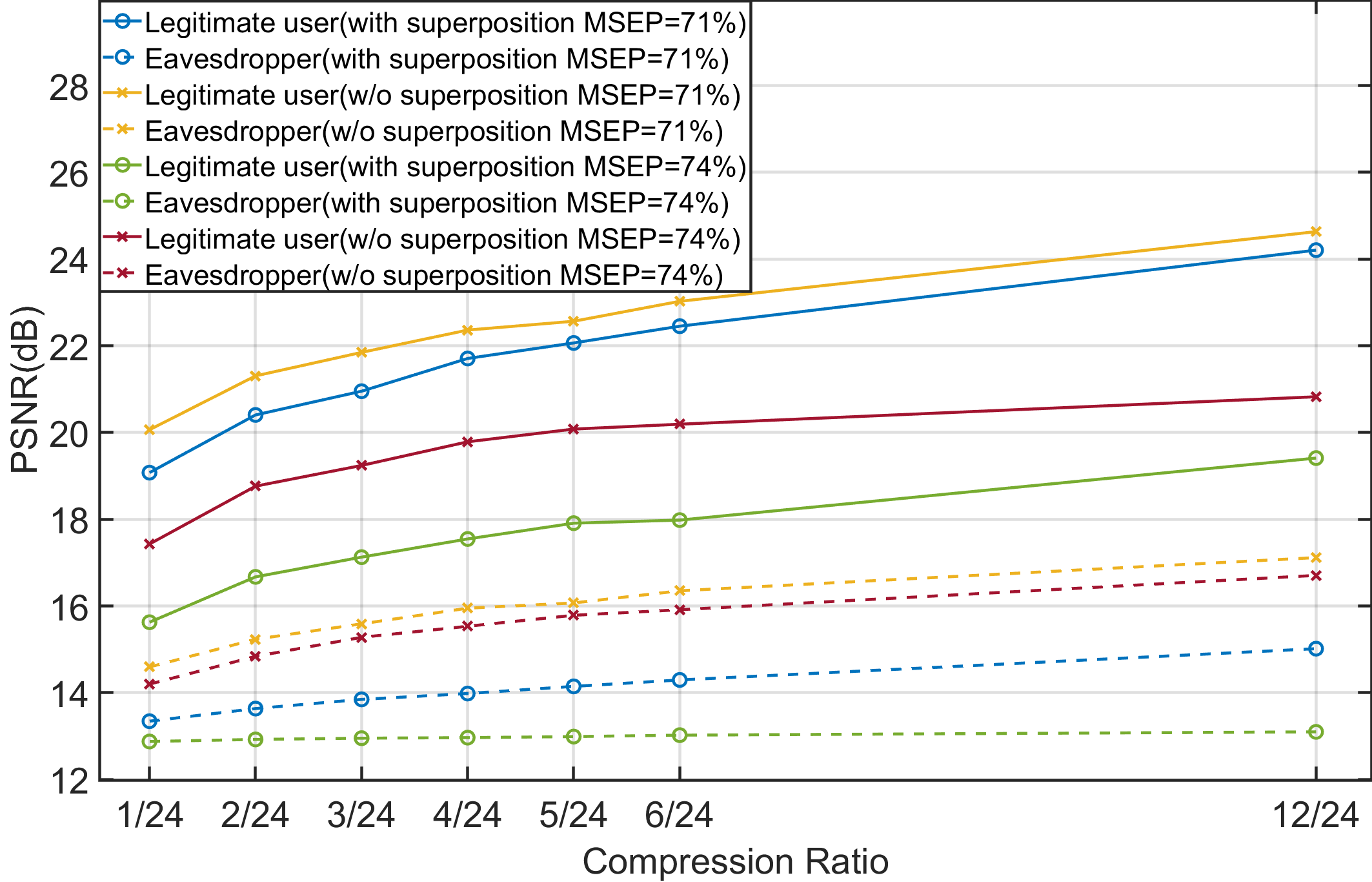}}
\caption{PSNR performance comparison of the proposed system with and without (w/o) superposition coding. Here, $\mathrm{SNR}_{\mathrm{leg}}=20\mathrm{dB}$ and $\mathrm{SNR}_{\mathrm{eve}}=-10\mathrm{dB}$. MSEP is set to 71\% or 74\%.}
\label{fig.plot_w_w_o_super}
\end{center}
\vskip -0.3in
\end{figure}

In this subsection, we analyze the impact of superposition coding on the proposed system's performance. We compare the PSNR performance of the legitimate user and the eavesdropper for our proposed system with and without (w/o) superposition coding, as shown in Fig.~\ref{fig.plot_w_w_o_super}. 
The channel conditions are set to $\mathrm{SNR}_{\mathrm{leg}}=20\mathrm{dB}$ and $\mathrm{SNR}_{\mathrm{eve}}=-10\mathrm{dB}$, 
with the MSEP set to 71\% or 74\%.
%
In the proposed system without superposition coding, the power of $\textbf{Y}_1$ remains unchanged, and $\textbf{Y}_2$ is not superposed with $\textbf{Y}_1$. 
%
%
As shown in Fig.~\ref{fig.plot_w_w_o_super}, when the jamming signal $\textbf{Y}_2$ is removed, both the legitimate user and the eavesdropper achieve higher PSNR performance. 
This indicates that superposition coding effectively protects the transmitted semantic information. 
Although the presence of $\textbf{Y}_2$ slightly degrades the performance of the legitimate user, our primary focus remains on system security. Therefore, the proposed method is beneficial for enhancing the security of SemCom systems.
In addition, it is worth noting that by deriving the SEPs for both the legitimate user and the eavesdropper under our superposition coding setting, we can explicitly control the system's security through PAC.
This indicates that our proposed method enables a controllable level of security for SemCom systems, allowing precise adjustment of the amount of noise added to the transmitted semantic information.

\section{Conclusion}
In this paper, we introduced a secure digital SemCom system based on superposition codes for wiretap channels. The proposed method involves generating a two-layered discrete superposition code through the overlay of one 4-QAM modulation constellation map onto another. 
Semantic information is modulated onto the outer constellation map, and a random constellation point is uniformly selected within the inner constellation. The power allocation between the inner and outer constellation maps is adjusted to achieve a high SEP for the eavesdropper, ensuring security and data privacy, while maintaining a low SEP for the legitimate user to facilitate data transmission. This method offers three significant advantages over conventional approaches.
Firstly, it enables data transmission that approaches the experimental upper bound of security by adjusting power allocation and setting the eavesdropper's SEP to the required level when both eavesdroppers and legitimate users utilize identical decoding schemes.
At a SEP that is high enough, the eavesdropper essentially decodes the outer constellation point as poorly as blinded random guessing. 
Secondly, the security performance of the proposed method remains robust even as the compression rate increases. This implies that we can enhance the legitimate user's performance by transmitting more data without concerns about information leakage. These advantages position our proposed method as a practical and secure SemCom solution, particularly suitable for scenarios with ample communication resources but stringent security requirements.
Thirdly, our proposed method enables quantifiable and controllable security, allowing for precise adjustments to the system's security level based on the specific requirements of the communication scenario. 
This level of flexibility is a key differentiator compared to existing methods, as most approaches do not provide explicit control over security.
Additionally, our proposed method does not rely on the assumption that the legitimate user has knowledge of the eavesdropper's network.

While the proposed system offers substantial security improvements, it still has limitations. A key challenge lies in the dependency on the channel SNR difference between the legitimate user and the eavesdropper. If the eavesdropper's channel quality improves significantly, the security of the system may be compromised. However, we hypothesize that by leveraging the inherent knowledge advantage of the legitimate user, such as incorporating additional training datasets or knowledge bases, it may be possible to mitigate this dependency. In such scenarios, the system's security could be enhanced further, potentially eliminating the reliance on the SNR disparity between the legitimate user and the eavesdropper. 
%
Looking ahead, our future efforts will focus on further improving the performance of the legitimate user. 
We believe that by encoding the inner constellation points with a random codeword from a codebook known only to the transmitter and legitimate user, rather than a totally random symbol sequence, the legitimate user can decode the energy jamming signal carried on the inner constellation points. 

%

\bibliographystyle{IEEEtran}

\bibliography{IEEEabrv,myref}

\begin{IEEEbiography}[{\includegraphics[width=1in,height=1.25in,clip,keepaspectratio]{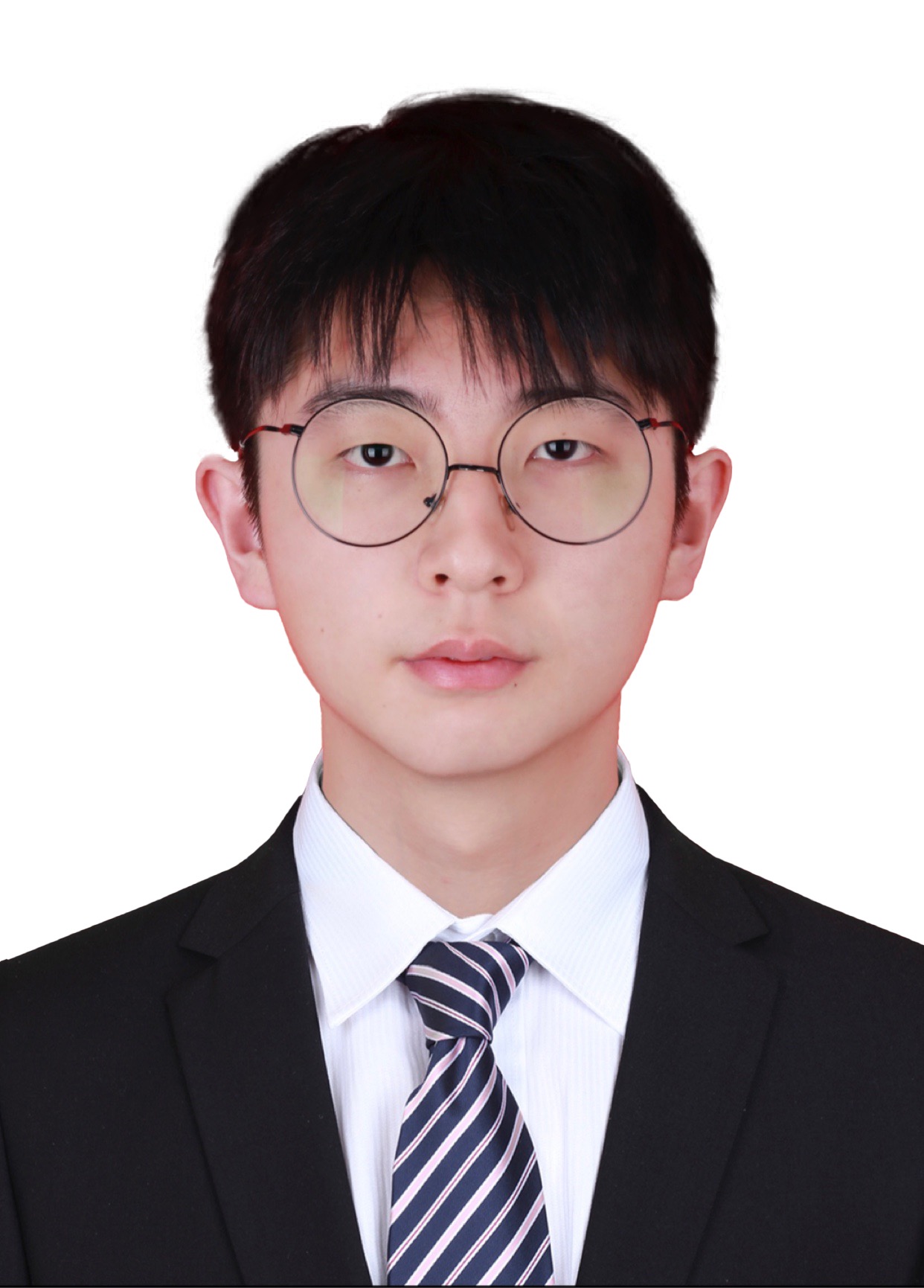}}]{Weixuan Chen}
(Graduate Student Member, IEEE) received the B.E. degree from Sichuan University, Chengdu, China, in 2022.
He is currently working toward the Ph.D. degree with the College of Information Science and Electronic Engineering, Zhejiang University, Hangzhou, China.
His research interests include semantic communications and computer vision.
\end{IEEEbiography}

\begin{IEEEbiography}[{\includegraphics[width=1in,height=1.25in,clip,keepaspectratio]{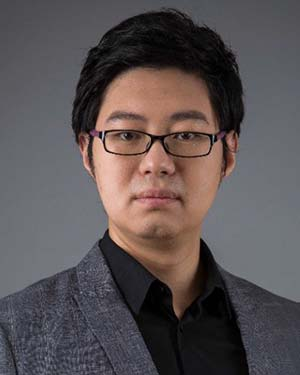}}]{Shuo Shao}
(Member, IEEE) received the B.S. degree in information science from Southeast University, Nanjing, China, in 2011, the M.A.Sc. degree in electrical and computer engineering from McMaster University, Hamilton, ON, Canada, in 2013, and the Ph.D. degree from Texas A\&M University, College Station, TX, USA, in 2017. He is currently an Associate Professor with the School of Electronics Information and Electrical Engineering, Shanghai Jiao Tong University, Shanghai, China. His research interests include network information theory, algebraic code, machine learning, and semantic communications.
\end{IEEEbiography}

\begin{IEEEbiography}[{\includegraphics[width=1in,height=1.25in,clip,keepaspectratio]{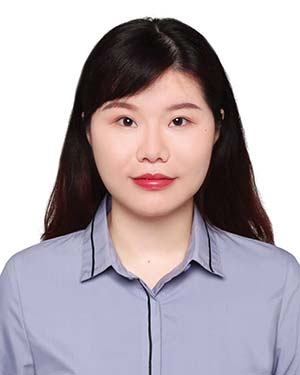}}]{Qianqian Yang}
(Member, IEEE) received the B.Sc. degree in automation from Chongqing University, Chongqing, China, in 2011, the M.S. degree in control engineering from Zhejiang University, Hangzhou, China, in 2014, and the Ph.D. degree in electrical and electronic engineering from Imperial College London, U.K. In 2016, she has held visiting positions with CentraleSupelec and was also with the New York University Tandon School of Engineering from 2017 to 2018. She was a Postdoctoral Research Associate with Imperial College London, and as a Machine Learning Researcher with Sensyne Health Plc. She is currently a Tenure-Tracked Professor with the Department of Information Science and Electronic Engineering, Zhejiang University. Her research interests mainly include wireless communications, information theory, and semantic communications. 
\end{IEEEbiography}

\begin{IEEEbiography}
[{\includegraphics[width=1in,height=1.25in,clip,keepaspectratio]{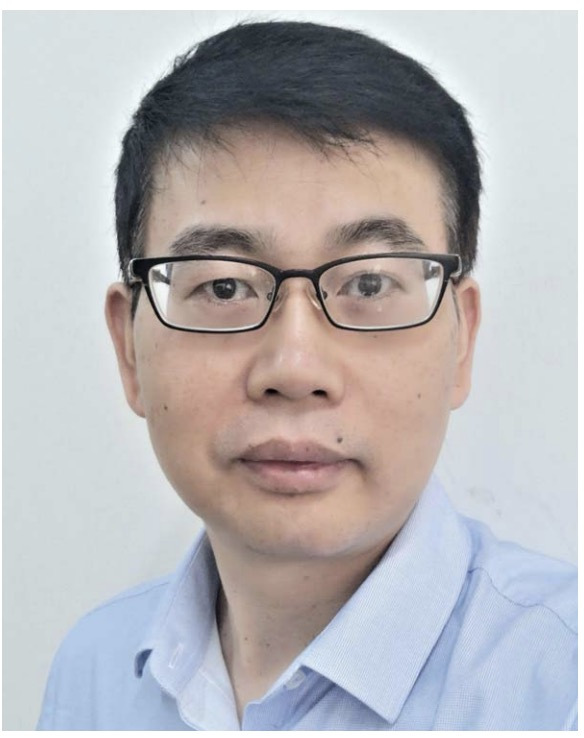}}]{Zhaoyang Zhang}
(Senior Member, IEEE) received the Ph.D degree from Zhejiang University, Hangzhou, China, in 1998. He is currently a Qiushi Distinguished Professor with Zhejiang University. He has coauthored more than 400 peer-reviewed international journals and conference papers, including eight conference best papers awarded by IEEE ICC 2019 and IEEE GLOBECOM 2020. His research interests include fundamental aspects of wireless communications and networking, such as information theory and coding theory, AI-empowered communications and networking, network signal processing and distributed learning, and synergetic sensing, computing and communication. He was awarded the National Natural Science Fund for Distinguished Young Scholar by NSFC in 2017. He is serving or has served as an Editor for IEEE Transactions on Wireless Communications, IEEE Transactions on Communications, and IET Communications, and as a General Chair, a TPC Co-Chair, or a Symposium Co-Chair for PIMRC 2021 Workshop on Native AI Empowered Wireless Networks, VTC-Spring 2017 Workshop on HMWC, WCSP 2013/2018, and GLOBECOM 2014 Wireless Communications Symposium. He was also a Keynote Speaker for GLOBECOM 2021 Workshop on Native-AI Wireless Networks, APCC 2018, and VTC-Fall 2017 Workshop NOMA.
\end{IEEEbiography}

\begin{IEEEbiography}[{\includegraphics[width=1in,height=1.25in,clip,keepaspectratio]{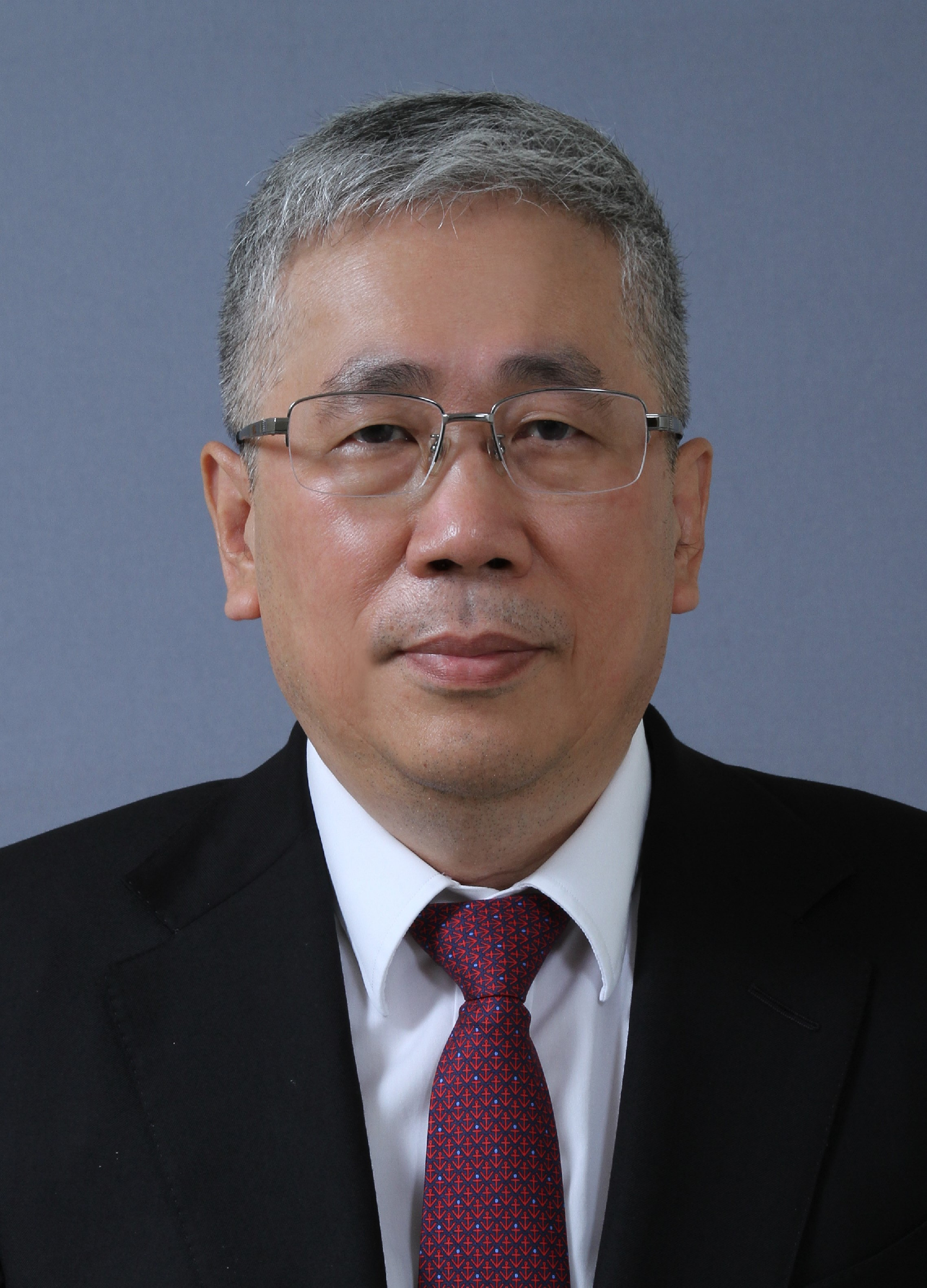}}]{Ping Zhang}
(Fellow, IEEE) received his M.S. degree in electrical engineering from Northwestern Polytechnical University, Xi'an, China, in 1986, and his Ph.D. degree in electric circuits and systems from BUPT, Beijing, China, in 1990. He is currently a Professor with BUPT, a Professor with the School of Information and Communication Engineering, Beijing University of Posts and Telecommunications (BUPT), and the Director of the State Key Laboratory of Networking and Switching Technology. He is also an Academician with the Chinese Academy of Engineering (CAE). His research interests mainly focus on wireless communications. He is also a member of the IMT-2020 (5G) Experts Panel and the Experts Panel for China's 6G development. He served as the Chief Scientist for the National Basic Research Program (973 Program), an Expert for the Information Technology Division of the National High-Tech Research and Development Program (863 Program), and a member of the Consultant Committee on International Cooperation, National Natural Science Foundation of China.
\end{IEEEbiography}
 
\vspace{12pt}
\end{CJK}
\end{document}